\documentclass[journal]{IEEEtran}
\usepackage{amsmath}
\usepackage{cite}
\usepackage{amssymb}
\usepackage{array}
\usepackage{amsfonts}
\usepackage{algorithm,algorithmic}
\usepackage{comment}
\usepackage{dsfont}
\usepackage{graphicx}
\usepackage{epsfig}
\usepackage{subfigure}
\usepackage{float}
\usepackage{epstopdf}
\usepackage{psfrag}
\usepackage{xcolor}
\usepackage{url}
\usepackage[colorlinks,linkcolor=black,urlcolor=black,anchorcolor=black,citecolor=black,hyperfootnotes=true]{hyperref}

\newtheorem{remark}{Remark}

\usepackage{mathtools}

\title{Joint Multi-User Tracking and Signal Detection in Reconfigurable Intelligent Surface-Assisted Cell-Free ISAC Systems}
\author{\IEEEauthorblockN{Weifeng~Zhu, 
Junyuan~Gao, 
Shuowen~Zhang, 
Meixia~Tao, 
and Liang~Liu}
\thanks{Weifeng Zhu, Junyuan Gao, Shuowen Zhang, and Liang Liu are with
Department of Electrical and Electronic Engineering, The Hong Kong Polytechnic University, Hong Kong SAR, China (email: \{eee-wf.zhu, junyuan.gao, shuowen.zhang, liang-eie.liu\}@polyu.edu.hk).}
\thanks{Meixia Tao is with
School of Information Science and Electronic Engineering, Shanghai Jiao Tong University, Shanghai 200240, China (email: mxtao@sjtu.edu.cn).}
}

\begin{document}
		\maketitle 

\abovedisplayskip=1pt
\belowdisplayskip=1pt
\allowdisplaybreaks

\begin{abstract}
This paper investigates the cell-free multi-user integrated sensing and communication (ISAC) system, where multiple base stations collaboratively track the users and detect their signals. Moreover, reconfigurable intelligent surfaces (RISs) are deployed to serve as additional reference nodes to overcome the line-of-sight blockage issue of mobile users for accomplishing seamless sensing. Due to the high-speed user mobility, the multi-user tracking and signal detection performance can be significantly deteriorated without elaborated online user kinematic state updating principles. To tackle this challenge, we first manage to establish a probabilistic signal model to comprehensively characterize the interdependencies among user states, transmit signals, and received signals during the tracking procedure. Based on the Bayesian problem formulation, we further propose a novel hybrid variational message passing (HVMP) algorithm to realize computationally efficient joint estimation of user states and transmit signals in an online manner, which integrates VMP and standard MP to derive the posterior probabilities of estimated variables. Furthermore, the Bayesian Cramér-Rao bound is provided to characterize the performance limit of the multi-user tracking problem, which is also utilized to optimize RIS phase profiles for tracking performance enhancement. Numerical results demonstrate that the proposed algorithm can significantly improve both tracking and signal detection performance over the representative Bayesian estimation counterparts.
\end{abstract} 


\section{Introduction}

The advent of the six-generation (6G) wireless networks marks a significant revolution to encompass advanced environmental sensing with information transmission under the common cellular infrastructure through integrated sensing and communication (ISAC) \cite{Liu_2022_CST,isac_survey1,isac_survey2}. Under the dual-functional integration, a variety of emerging intelligent applications including autonomous driving, low-altitude economy, and smart manufacturing \cite{isac_survey1,isac_survey2,Liu_2024_CM} can be supported by 6G cellular systems without embedding extra infrastructures. 
As the basic issue in ISAC, the acquisition of accurate position information is usually realized by measuring the line-of-sight (LOS) channel parameters towards the BS \cite{Dong_2012_TSP,Yuan_2021_JSTSP,Zhu_2025_TWC}, which includes the delay, Doppler frequency, angle of arrival (AOA), etc. However, the LOS path towards the BS can often be blocked by various obstacles in practical scenarios, leading to missing position measurements.


To overcome the reliance on the LOS channel existence, reconfigurable intelligent surface (RIS) emerges as a promising technique in ISAC systems by reshaping electromagnetic propagation environments \cite{Tang_2021_TWC,Wu_2021_TCOM,Liu_2021_CST,Zhu_2025_WCSP}. In particular, RISs can be widely deployed to provide seamless surveillance coverage by creating virtual LOS channels along the user-RIS-BS paths, which bypass environment obstacles via controllable reflections. By regarding RISs as additional anchors, our preliminary work \cite{Zhu_2026_TWC} rigourously show the feasibility to acquire the position information from these virtual LOS channels, when the LOS blockage occurs. 
On the other hand, RISs are also capable of mitigating the wireless interference and fading issue under advanced phase profile design, which significantly enhances the communication quality \cite{Tang_2021_TWC,Wu_2021_TCOM}. 
Thanks to these attractive benefits, the RIS becomes one of the indispensable basic hardware facilities in 6G ISAC systems \cite{Chepuri_2023_SPM}.

In the context of RIS-assisted ISAC, early researches mainly focus on the perspectives of theoretical performance analysis and optimization, where the sensing performance is characterized by sensing signal-to-noise ratio (SNR) \cite{liu_2022_JSTSP} and Cramér-Rao bound (CRB) \cite{Elz_2021_TSP,Song_2023_TSP,Zhang_2025_CL}. 
From the perspective of algorithm design, the work \cite{Zhang_2021_TWC} proposes a fingerprint-based method to localize indoor users based on their unique received signal strength (RSS) patterns. However, the high-resolution RSS information is hard to be obtained in the outdoor scenario, which restricts the applications. Alternatively, the work \cite{Wang_2024_TWC} proposes a trilateration-based method by utilizing the range information from both direct and reflecting links, while the angular measurement capability of RISs is not excavated with limited localization resolution. Then, the works \cite{Han_2022_JSTSP,Rahal_2024_JSTSP} leverage the maximum likelihood methods to estimate both of the distance and AOA towards the RIS for localization, while their complexities are prohibitively intensive in the multi-user scenario. In our work \cite{Zhu_2026_TWC}, an efficient subspace-based algorithm and Discrete Fourier Transform (DFT)-based RIS phase profile are designed to facilitate the multi-user localization problem. However, these works perform localization in each time slot independently, which neglects the temporal correlation of user kinematic states in adjacent frames. As such, in the mobility case, these methods may only provide limited estimation performance on position information.

To promote high-accuracy position tracking of mobile users, it is beneficial to leverage both of the historical and fresh observations to acquire the position information in each time slot. Following this guideline, recent works \cite{Ammous_2022_VTC,Zheng_2024_JSAC} propose to apply the famous Kalman filter (KF) technique \cite{Einicke_1999_TSP} to improve the tracking performance. In \cite{Ammous_2022_VTC}, an extended KF method is proposed to track the mobile user based on the range measurements towards the BS and RIS. Then, the work \cite{Zheng_2024_JSAC} proposes a manifold-based unscented KF method for user tracking based on delay-Doppler-angle measurements. However, these two works are limited to the single-user case and assume that the LOS path to the BS always exists. Alternatively, the work \cite{Teng_2023_JSAC} proposes a Bayeisan tracking algorithm for the multi-user tracking problem in the pure LOS blockage scenario based on only the position-related measurements towards the RIS. In practice, the blockage conditions of direct and reflecting links are usually random and unknown, implying that the received signals contain a combination of measurements from these two types of links. 

In contrast to the aforementioned works, this work aims to provide a comprehensive investigation on the general RIS-assisted cell-free ISAC system with unknown link blockages, which has the following three differences. First, it is necessary to identify the blockage conditions of both direct and reflecting links in each time slot, which then enables us to extract and combine the kinematic measurements in received signals to accomplish the tracking task.
Second, compared with single-BS systems \cite{Ammous_2022_VTC,Zheng_2024_JSAC,Teng_2023_JSAC}, there exist strong correlations among the reflecting links to all BSs through the identical RIS in cell-free ISAC systems, which share the same position-related parameters towards the RIS. Thus, this work proposes to directly estimate the position-related parameters towards each RIS from all the related reflecting links instead of treating each reflecting link individually.
Third, note that the above works \cite{Zhang_2021_TWC,Wang_2024_TWC,Han_2022_JSTSP,Rahal_2024_JSTSP,Zhu_2026_TWC,Ammous_2022_VTC,Zheng_2024_JSAC,Teng_2023_JSAC} all apply the pilot signals to the sensing task, leading to significantly increased sensing overhead in the multi-user scenario. To address this issue, we are motivated to leverage the data signals to facilitate the sensing task for boosting both of the sensing and communication performance. 

In this work, we studies the joint multi-user tracking and signal detection problem in the RIS-assisted cell-free ISAC system. 
The main distinctions and contributions of this work are summarized as follows:
\begin{enumerate}
  \item For the considered system, we first propose an ISAC uplink transmission protocol for realizing joint multi-user tracking and communication. Then, we establish a novel probabilistic signal model that includes geometric constraints and statistical dependencies in the received signals for the considered system. Subsequently, we formulate the joint multi-user tracking and signal detection problem under the Bayesian inference framework, enabling to realize either \emph{maximum a posterior} (MAP) or \emph{minimum mean square error} (MMSE) estimation.
  \item Under the graphical representation of the probabilistic model, we devise a novel hybrid variational message passing (HVMP) algorithm for the joint estimation of user states and transmit signals. Specifically, the HVMP algorithm integrates VMP and MP to realize efficient calculations of the desired posterior probabilities and side information for future tracking, both of which are given in explicit expressions. Compared with existing works, the blockage condition of direct and reflecting links can also be automatically estimated, and the position-related measurements of all available links can be derived and flexibly combined to facilitate the joint estimation. 
  \item We provide the Bayesian Cramér-Rao bound (BCRB) to characterize the performance limit of the considered problem. Meanwhile, we formulate a weighted BCRB minimization problem to optimize the RIS phase profile, which enhances the tracking and detection performance over the traditional heuristic RIS phase profile design.
  \item Extensive numerical results are provided to validate the performance superiority of the proposed approach to representative benchmarks. It is verified that the HVMP can provide consistently centimeter-level tracking accuracy even if there are a large number of tracking users. Moreover, the millimeter-level tracking accuracy can potentially be accomplished by effectively combining measurements from both direct and reflecting links when the blockage condition becomes less serious.
\end{enumerate}


The rest of the paper is organized as follows. Section II introduces the system model of the considered RIS-assisted ISAC system and Section III provides the Bayesian problem formulation of the considered problem. In Section IV, the proposed HVMP algorithm is developed, whose message passing equations are specified. In Section V, the BCRB is obtained for performance analysis and a RIS phase profile optimization problem is studied. Finally, Section VI evaluates the performance of the proposed multi-user tracking and signal detection algorithm and Section VII concludes this work.

\emph{Notations:}
In this paper, vectors and matrices are denoted by boldface lower-case letters and boldface upper-case letters, respectively. For a complex matrix $\mathbf{X}$, $[\mathbf{X}]_{\mathcal{N},\mathcal{M}}$ denote the sub-matrix with rows and columns in $\mathcal{N}$ and $\mathcal{M}$, respectively, while $\mathbf{X}^\mathsf{T}$ and $\mathbf{X}^\mathsf{H}$ denote its transpose, and conjugate transpose, respectively. For a complex number $x \in \mathbb{C}$, its phase is denoted as $\angle x$. Further, $\mathbb{E}_{\mathbf{a}}[\cdot]$ denotes the expectation operation over random vector $\mathbf{a}$, $\odot$ denotes the Hadamard product, and $\otimes$ denotes the Kronecker product. 
Gaussian distributions over the real random vector $\mathbf{x} \in \mathbb{R}^{N \times 1}$ and complex random vector $\mathbf{y} \in \mathbb{C}^{N \times 1}$ are denoted by $\mathcal{N}(\mathbf{x};\bar{\mathbf{x}},\pmb{\Sigma}_{\mathbf{x}})$ and $\mathcal{CN}(\mathbf{y};\bar{\mathbf{y}},\pmb{\Sigma}_{\mathbf{y}})$, respectively.
The von Misés (VM) distribution over the variable $\theta$ is denoted by  $\mathcal{VM}(\theta;\mu,\kappa) = \left(2\pi I_0(\kappa)\right)^{-1} \exp(\kappa \cos(\theta - \mu))$, with $I_0(\kappa)$ denoting the zeroth-order Bessel function.

\section{System Model}\label{sec:system_model}

We consider an uplink cell-free ISAC system with $G$ BSs each with $M_{\mathsf{B}}$ antennas, $K$ single-antenna mobile users, and $R$ RISs each with $M_{\mathsf{I}}$ reflecting elements, as shown in Fig. 1. 
In the two-dimension (2D) system, the coordinate of each BS $g \in \mathcal{G} =  \{1,2,\dots,G\}$ is denoted as $\mathbf{p}^{\mathsf{B}}_{g} = [p^{\mathsf{B}}_{x,g},p^{\mathsf{B}}_{y,g}]^\mathsf{T} \in \mathbb{R}^{2 \times 1}$ and the coordinate of each RIS $r \in \mathcal{R} =  \{1,2,\dots,R\}$ is denoted as $\mathbf{p}^{\mathsf{I}}_{r} = [p^{\mathsf{I}}_{x,r},p^{\mathsf{I}}_{y,r}]^\mathsf{T} \in \mathbb{R}^{2 \times 1}$. In each time slot $t$, the position and velocity of each user $k$ are denoted as $\mathbf{p}^{\mathsf{U}}_{k,t} = [p^{\mathsf{U}}_{x,k,t},p^{\mathsf{U}}_{y,k,t}]^\mathsf{T} \in \mathbb{R}^{2 \times 1}$ and $\dot{\mathbf{p}}^{\mathsf{U}}_{k,t} = [\dot{p}^{\mathsf{U}}_{x,k,t},\dot{p}^{\mathsf{U}}_{y,k,t}]^\mathsf{T} \in \mathbb{R}^{2 \times 1}$, respectively. 
All the users keep transmitting orthogonal frequency division multiplexing (OFDM) signals in the uplink for $V$ time slots, while the BSs need to dynamically track the users and detect their transmit signals simultaneously. 

\begin{figure}
  \centering
  \includegraphics[width=.45\textwidth]{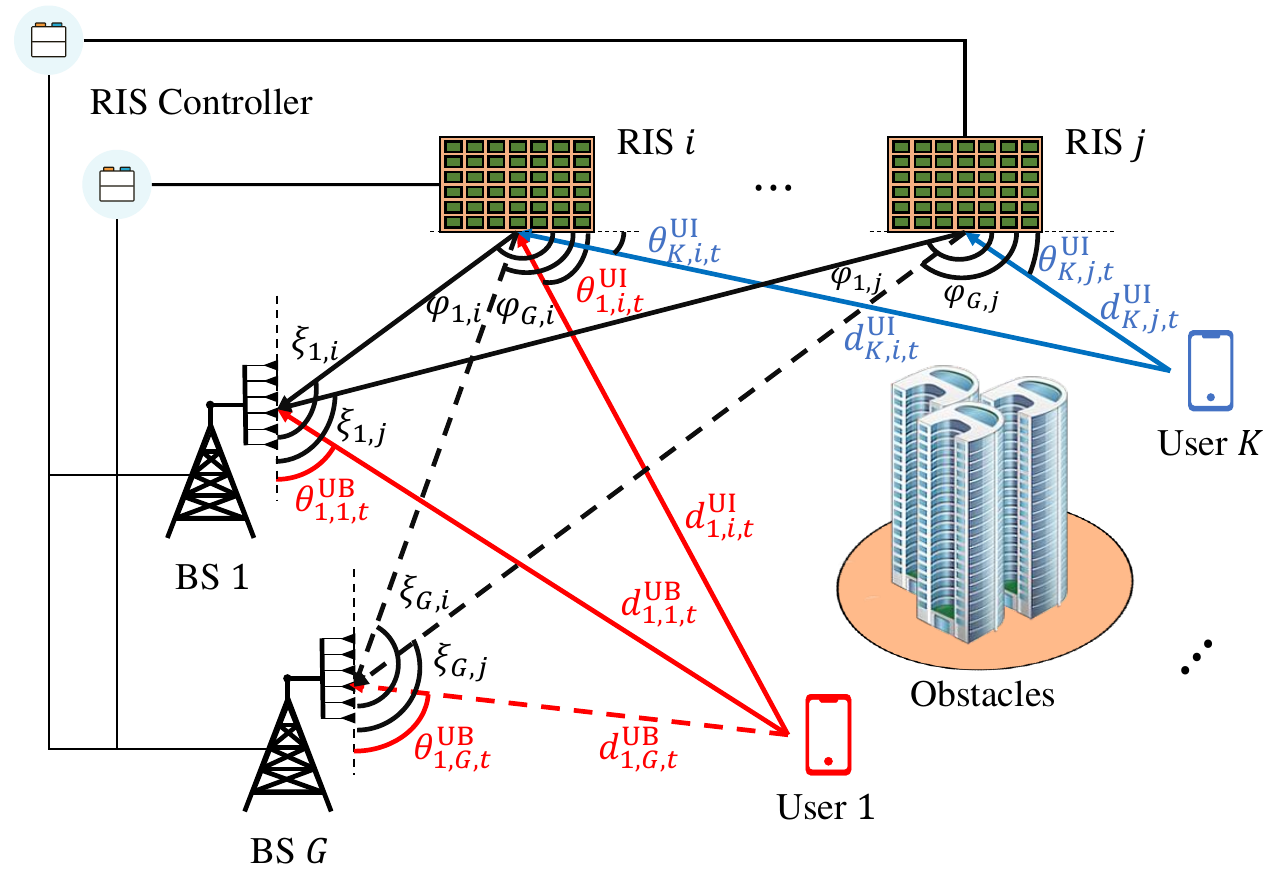}
  \vspace{-0.5cm}
  \caption{Illustration of the system model. Multiple RISs are deployed in the system to assist the sensing and communication service of users.}\label{fig:system model}
  \vspace{-0.5cm}
\end{figure}

\subsection{Mobility Model}\label{subsec:state model}

We assume that the motion of each user follows the discrete-time state model \cite{Tichavsky_1998_TSP}.
Define the state vector for all users at each time slot $t$ as $\pmb{\psi}_t = [\pmb{\psi}_{1,t}^\mathsf{T}, \dots, \pmb{\psi}_{K,t}^\mathsf{T}]^\mathsf{T} \in \mathbb{R}^{4K \times 1}$ with $\pmb{\psi}_{k,t} = [(\mathbf{p}_{k,t}^{\mathsf{U}})^\mathsf{T}, (\mathbf{\dot{p}}_{k,t}^{\mathsf{U}})^\mathsf{T}]^\mathsf{T} \in \mathbb{R}^{4 \times 1}$. As such, the state transition equation \cite{Tichavsky_1998_TSP,Djuric_2003_SPM,Bar_2009_CSM}
can be written by\footnote{In practice, the velocity of each user $k$ usually cannot keep perfectly constant in the whole time slot $t$. To address this issue, the velocity uncertainty is modeled as the motion noise term in the discrete-time state model.}
\begin{align}\label{equ:discrete_mm}
    \pmb{\psi}_{t} = \mathbf{F} \pmb{\psi}_{t-1} + \pmb{\omega}_{t-1}, ~\forall t,
\end{align}
where the transition matrix $\mathbf{F}$ is defined as $\mathbf{F} = \mathbf{I}_{K} \otimes \mathbf{F}_0$ with
\begin{align}
    \mathbf{F}_0 &= \begin{bmatrix}
        \mathbf{I}_{2} & \Delta T \mathbf{I}_{2}  \\
        \mathbf{0} & \mathbf{I}_{2} 
    \end{bmatrix} \in \mathbb{R}^{4K \times 4K}.
\end{align}
In the mobility model \eqref{equ:discrete_mm}, $\pmb{\omega}_{t-1} \in \mathbb{R}^{4K \times 1}$ denotes the motion noise following the Gaussian distribution $\mathcal{N}(\mathbf{0},\mathbf{Q}_{t-1})$ with $\mathbf{Q}_{t-1} = \text{blkdiag}(\mathbf{Q}_{1,t-1},\dots,\mathbf{Q}_{K,t-1})$, and $\Delta T$ denote the time interval between two adjacent tracking time slots. 
It is assumed that the motion noise vectors $\mathbf{w}_t$ and $\mathbf{w}_{t'}$ are mutually independent for $t \ne t'$. 
From the above mobility model, we can find that the historical user state measurements can be leveraged to help refine the multi-user tracking performance, which then enhances the signal detection performance as well.

\subsection{Signal Model}

In the above system, each user transmits 
$Q$ OFDM symbols at each time slot, while the number of sub-carriers is $N$ and the sub-carrier spacing is $\Delta f$ Hz, such that the bandwidth is $B = N\Delta f$ Hz. Thus, there are totally $QN$ resource elements in the uplink transmission phase, contributing to the resource element set $\mathcal{E} = \{(q,n)| q = 1,\dots,Q, n = 1,\dots,N\}$. To avoid the inter-symbol interference, a cyclic prefix (CP) comprised by $J$ OFDM samples is inserted at the beginning of each transmit OFDM symbol. As such, the period of each OFDM symbol is $\Delta t = \frac{N+J}{N \Delta f}$. 

At each BS $g$, the received signal at the $(q,n)$-th resource element in each time slot $t$ can be given by
\begin{align}\label{equ:spatial_signal_gnqt}
    \mathbf{y}^{(q)}_{g,n,t} &= \sum_{k=1}^{K} s^{(q)}_{k,n,t} \mathbf{h}^{(q)}_{k,g,n,t} + \mathbf{z}^{(q)}_{g,n,t}, ~\forall g,n,q,t,
\end{align} 
where $s^{(q)}_{k,n,t}$ is the data signal at the $(q,n)$-th resource element transmitted from the user $k$, $\mathbf{h}^{(q)}_{k,g,n,t}  \in \mathbb{C}^{M_{\mathsf{B}} \times 1}$ is the channel between the user $k$ and the BS $g$ at the $(q,n)$-th resource element, and $\mathbf{z}^{(q)}_{g,n,t} \in \mathbb{C}^{M_{\mathsf{B}} \times 1}$ is the background noise vector following complex Gaussian distribution of $\mathcal{CN}(\mathbf{z}^{(q)}_{g,n,t};0,\sigma^2_z \mathbf{I})$.

\subsection{Channel Model}

In the signal model \eqref{equ:spatial_signal_gnqt}, each of the effective channel vectors, i.e., $\{\mathbf{h}^{(q)}_{k,g,n,t}\}$, is comprised by the direct user-BS channel and cascaded user-RIS-BS channels. Specifically, for the $(q,n)$-th resource element at time slot $t$,  let $\mathbf{h}_{k,g,n,t}^{{\mathsf{UB}}, (q)}$, $\mathbf{h}_{k,r,n,t}^{{\mathsf{UI}}, (q)}$, and $\mathbf{H}^{\mathsf{IB}}_{g,r,n}$ denote the channels from user $k$ to the BS $g$, from user $k$ to the RIS $r$, and from the RIS $r$ to the BS $g$, respectively. Then, the corresponding channel vector $\mathbf{h}^{(q)}_{k,g,n,t}$ is denoted as
\begin{align}\label{equ:UE-BS_channel}
  \mathbf{h}_{k,g,n,t}^{(q)} 
  &=  \mathbf{h}_{k,g,n,t}^{{\mathsf{UB}}, (q)} + \sum_{r=1}^{R} \mathbf{H}^{\mathsf{IB}}_{g,r,n} \text{diag}(\pmb{\phi}_{r,t}^{(q)}) \mathbf{h}_{k,r,n,t}^{{\mathsf{UI}}, (q)}. 
\end{align}
where $\pmb{\phi}_{r,t}^{(q)}=[\phi_{1,r,t}^{(q)},\cdots,\phi_{M_\mathsf{I},r,t}^{(q)}]^\mathsf{T} \in \mathbb{C}^{M_\mathsf{I}\times 1}$ is the phase profile vector of the RIS $r$ during the $q$-th OFDM symbol duration in the time slot $t$ with $|\phi_{m_\mathsf{I},r,t}^{(q)}| = 1, ~\forall m_{\mathsf{I}},r,t,q$. 
In this work, we assume that the user-BS channels, user-RIS channels, and RIS-BS channels all follow the binary LOS channel model\footnote{The small-scale fading component generally has much weaker power than the LOS path (smaller than -20 dB) \cite{Akdeniz_2014_JSAC,Han_2015_TWC} and thus can be eliminiated, due to severe path loss and reflection loss.}, under which each channel is either a LOS channel in the case without blockage or zero in the case with blockage.
Specifically, the RIS $r$-BS $g$, user $k$-BS $g$, and user $k$-RIS $r$ channels can be given by
\begin{align}
  \mathbf{H}^{\mathsf{IB}}_{g,r,n} =&~ \alpha^{\mathsf{IB}}_{g,r}\beta^{\mathsf{IB}}_{g,r} e^{-\jmath \zeta_{\mathsf{F}} \tau^{\mathsf{IB}}_{g,r}} \mathbf{a}_{\mathsf{B}}(\theta^{\mathsf{B}}_{g,r}) \mathbf{a}_{\mathsf{I}}^\mathsf{T}(\theta^{\mathsf{I}}_{g,r}), \label{equ:h_IB} \\
  \mathbf{h}_{k,g,n,t}^{{\mathsf{UB}}, (q)} =&~  \alpha_{k,g,t}^{\mathsf{UB}} \beta^{\mathsf{UB}}_{k,g,t} e^{-\jmath (\zeta_{\mathsf{F}} (n-1) \tau^{\mathsf{UB}}_{k,g,t} + \zeta_{\mathsf{T}} (q-1) \nu_{k,g,t}^{\mathsf{UB}})} \notag \\ 
  &\times \mathbf{a}_{\mathsf{B}}(\theta^{\mathsf{UB}}_{k,g,t}), \label{equ:h_UB} \\ 
  \mathbf{h}_{k,r,n,t}^{{\mathsf{UI}}, (q)} =&~  \alpha_{k,r,t}^{\mathsf{UI}} \beta^{\mathsf{UI}}_{k,r,t} e^{-\jmath (\zeta_{\mathsf{F}} (n-1) \tau^{\mathsf{UI}}_{k,r,t} + \zeta_{\mathsf{T}} (q-1) \nu_{k,r,t}^{\mathsf{UI}})} \notag \\ 
  &\times \mathbf{a}_{\mathsf{I}}(\theta^{\mathsf{UI}}_{k,r,t}), \label{equ:h_UI} 
\end{align}
respectively, where $\zeta_{\mathsf{F}} = 2 \pi \Delta f$ and  $\zeta_{\mathsf{T}} = 2 \pi \Delta t$.
In the above, $\alpha \in \{0,1\}$, $\beta$, $\tau$, $\nu$, and $\theta$ are the blockage indicator, complex channel gain, delay profile, AOA (angle of departure (AOD)), respectively, whose superscripts and subscripts are omitted for clarity. The geometric relationship between the position-related parameters ($\tau$, $\nu$, and $\theta$) and user states is detailed provided in Section \ref{subsec:prob_SR}. Consider that the uniform linear array (ULA) with $d$ spacing is employed for both BSs and RISs, the steering vectors $\mathbf{a}_{\mathsf{i}}(\theta), ~\forall \mathsf{i} \in \{\mathsf{B}, \mathsf{I}\}$ with the AOA (AOD) $\theta$ can be expressed by
\begin{align}\label{equ:steer_vec}
    \mathbf{a}_{\mathsf{i}}(\theta) = [ 1, e^{-\jmath \zeta_{\mathsf{S}} \cos\theta}, \dots, e^{-\jmath \zeta_{\mathsf{S}} (M_{\mathsf{i}}-1) \cos\theta} ]^\mathsf{T} \in \mathbb{C}^{M_{\mathsf{i}} \times 1}, 
\end{align}
where $\zeta_{\mathsf{S}} = \frac{2\pi d}{\lambda}$.

Based on the mobility model \eqref{equ:discrete_mm}, signal model \eqref{equ:spatial_signal_gnqt}, and the above channel model, the BSs aim to achieve two goals: tracking the user positions, i.e., estimating $\{\mathbf{p}^{\mathsf{U}}_{k,t}\}_{k=1}^{K}, ~\forall t$, and decoding user messages, i.e., estimating $s^{(q)}_{k,n,t}, ~\forall k,n,q,t$.  

\begin{remark}
Note that there are quite a lot of works considering localization and tracking problem in wireless networks \cite{Han_2025_TWC,Dong_2012_TSP,Yuan_2021_JSTSP}. Compared to these works, we have two new challenges. First, we consider the case when the LOS channels for user-BS and user-RIS links can be blocked, due to various obstacles. Thus, it is also essential to estimate the blockage conditions of all LOS channels during the tracking procedure, which enables to extract the user state information. Second, in \cite{Han_2025_TWC,Dong_2012_TSP,Yuan_2021_JSTSP,Teng_2023_JSAC}, each user transmits known pilot signals to the BS, which only needs to focus on the tracking job. On the contrary, in our considered system, the BS can utilize the data signal to perform tracking. Therefore, the BSs need to estimate both the position-related information (angle, delay, and Doppler frequency) and the unknown user messages simultaneously from their received signals. The above two new challenges call for totally new signal processing solutions in our setup.
\end{remark}



\section{Problem Formulation}


In this section, we first introduce an ISAC protocol such that BSs can simultaneously track the users and detect their signals based on their received signals. Then, we formulate the joint tracking and signal detection problem under this protocol. 

\subsection{ISAC Protocol}

\begin{figure}[t]
  \centering
  \subfigure[Transmission signal structure for uplink users.]{
  \includegraphics[width=.47\textwidth]{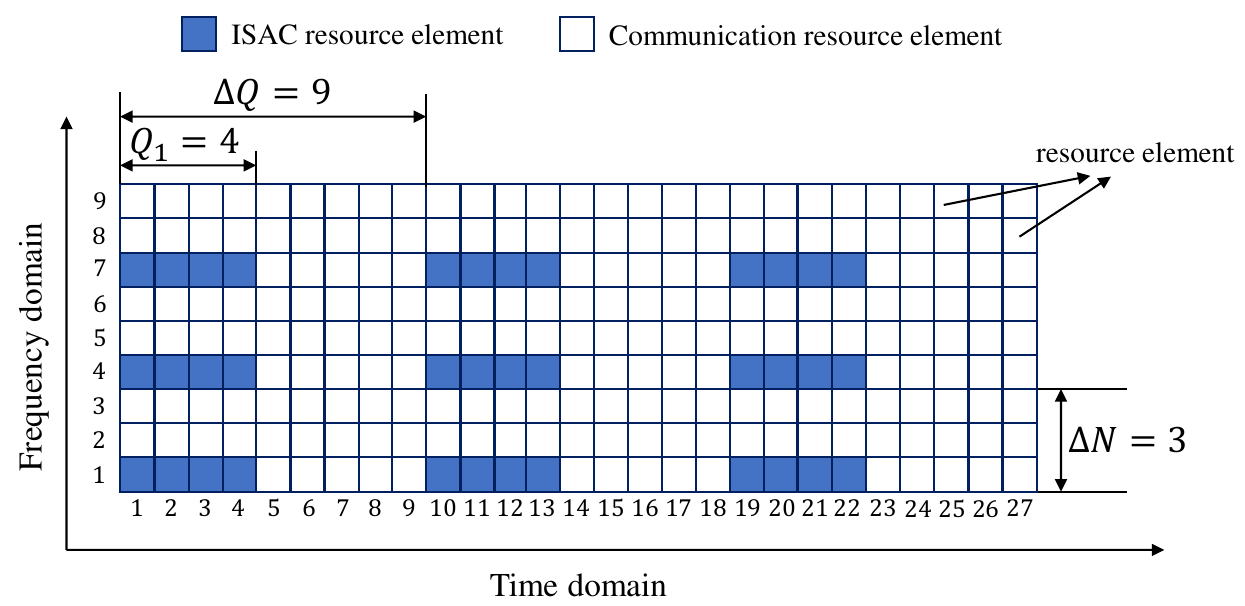}}
  \subfigure[Reflection pattern of the phase profile for each RIS.]{
  \includegraphics[width=.47\textwidth]{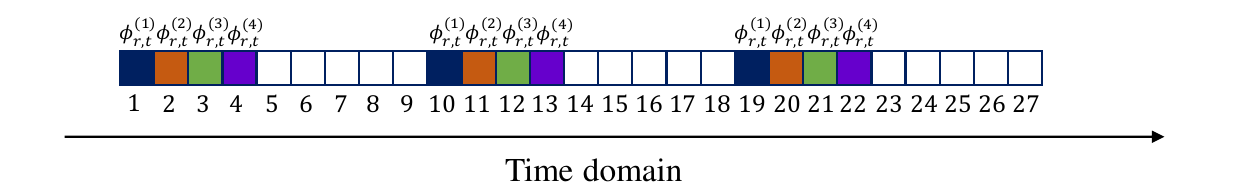}}
  \vspace{-0.1cm}
  \caption{An example illustration of the ISAC protocol with $Q_1 = 4$, $I = 3$, $\Delta Q = 9$, and $\Delta N = 3$. Then, we have $\mathcal{N}_{\rm I} = \{1,4,7\}$ and $\mathcal{Q}_{\rm I} = \{ 1,2,3,4 \} \cup \{ 10,11,12,13 \} \cup \{ 19,20,21,22 \}$.}\label{fig:frame}
  \vspace{-0.5cm}
\end{figure}

Under the proposed protocol, all the resource elements are divided into two sets, i.e., Set I, which is termed as the ISAC set $\mathcal{E}_{\rm I}$ with $E_{\rm I} = Q_{\rm I}N_{\rm I}$ resource elements, and Set II, which is termed as the communication set $\mathcal{E}_{\rm C} = \mathcal{E} \setminus \mathcal{E}_{\rm I}$ with $E_{\rm C} =  (QN - Q_{\rm I}N_{\rm I}$) resource elements. Specifically, all the BSs leverage received signals in the ISAC Set to jointly track the users and detect their signals at each time slot $t$. Then, since the user channels are known with their estimated AOA/AOD, delay, and Doppler information, BSs only need to detect the signals in the communication set. In the rest of this paper, we merely focus on the ISAC set, because signal detection with known channels in the communication set is standard. 

In this work, we consider the ISAC set given by $\mathcal{E}_{\rm I} = \{(q,n) | q \in \mathcal{Q}_{0}, n \in \mathcal{N}_{0}\}$, which has a comb-like pattern shown in Fig. \ref{fig:frame}(a) to facilitate precise estimation on both delays and Doppler frequencies. Specifically, we have $\mathcal{N}_0 = \{\varsigma_{\breve{n}} | \varsigma_{\breve{n}} =  1+ (\breve{n}-1)\Delta N, \breve{n}=1,\dots,N_{\rm I}\}$ and $\mathcal{Q}_0 = \bigcup_{i = 1}^{I} \mathcal{Q}_i$, respectively, where $\Delta N \ge 1$ and $\mathcal{Q}_i = \{\iota_{i,\breve{q}} | \iota_{i,\breve{q}} = \breve{q}+(i-1)\Delta Q, \breve{q}=1,\dots,Q_1 \}$, $\forall i$ with $Q_{\rm I} = I Q_1$, $Q_1 > 1$, and $\Delta Q \gg 1$.\footnote{The spacing $\Delta Q$ is usually set to be large enough to facilitate the Doppler frequency estimation.}

Next, we introduce the user transmission strategy in the ISAC set. Similar to \cite{Nassaji_2022_TCOM,Vem_2019_TCOM}, we adopt repetition coding at the user side. Specifically, we have 
\begin{align}
    s_{k,n,t}^{(q)} &= \tilde{s}_{k,t}, ~\forall k,t, (q,n) \in \mathcal{E}_{\rm I}.
\end{align}

The reason to adopt repetition coding in the ISAC set is as follows. From the model \eqref{equ:spatial_signal_gnqt} -- \eqref{equ:h_UI}, we require to estimate $E_{\rm I}$ independent signals and one delay-Doppler pair from $E_{\rm I}$ signal measurements with independent coding, which is generally an underdetermined problem and intractable. Thus, repetition coding is motivated to address this issue by reducing the number of estimated signals, contributing to an overdetermined problem.
Note that repetition coding may reduce the spectral efficiency (SE). However, this SE loss is inevitable if BSs have to jointly track the users and detect their signals. Moreover, repetition coding is merely used in the ISAC set, while we do not need repetition coding anymore for the communication set since the channels are known for signal detection. 

Moreover, for extracting the AOA information of user-RIS links, we follow our preliminary work \cite{Zhu_2026_TWC} to design the RIS phase profile with $\pmb{\phi}_{r,t}^{(\iota_{i,\breve{q}})} = \tilde{\pmb{\phi}}_{r,t}^{(\breve{q})}$, $\forall r,t,i,\breve{q} = 1,\dots,Q_1$ and $\tilde{\pmb{\phi}}_{r,t}^{(\breve{q}_1)} \ne \tilde{\pmb{\phi}}_{r,t}^{(\breve{q}_2)}, ~\forall \breve{q}_1 \ne \breve{q}_2$. As shown in Fig. \ref{fig:frame}(b), this design implies that the RIS has the common RIS phase pattern in each subset $\mathcal{Q}_i, \forall i$.  
This also explains why we consider $Q_1 > 1$ for each $\mathcal{Q}_i$, since the AOA of each user-RIS link can never be extracted with $Q_1 = 1$ \cite{Zhu_2026_TWC}.

\begin{remark}
Note that the proposed ISAC protocol differs from the traditional OFDM transmission protocol only in terms of transmit signals in the ISAC sub-block. Specifically, we replace the pilot signals in the traditional OFDM protocol with information signals to enhance the spectral efficiency. This also means that the ISAC protocol can be easily applied to existing communication systems. In Section \ref{sec:simu}, we will show that the proposed ISAC protocol has the approaching tracking performance to the pilot-based transmission paradigm.
\end{remark}

\begin{figure}[t]
  \centering
  \includegraphics[width=.48\textwidth]{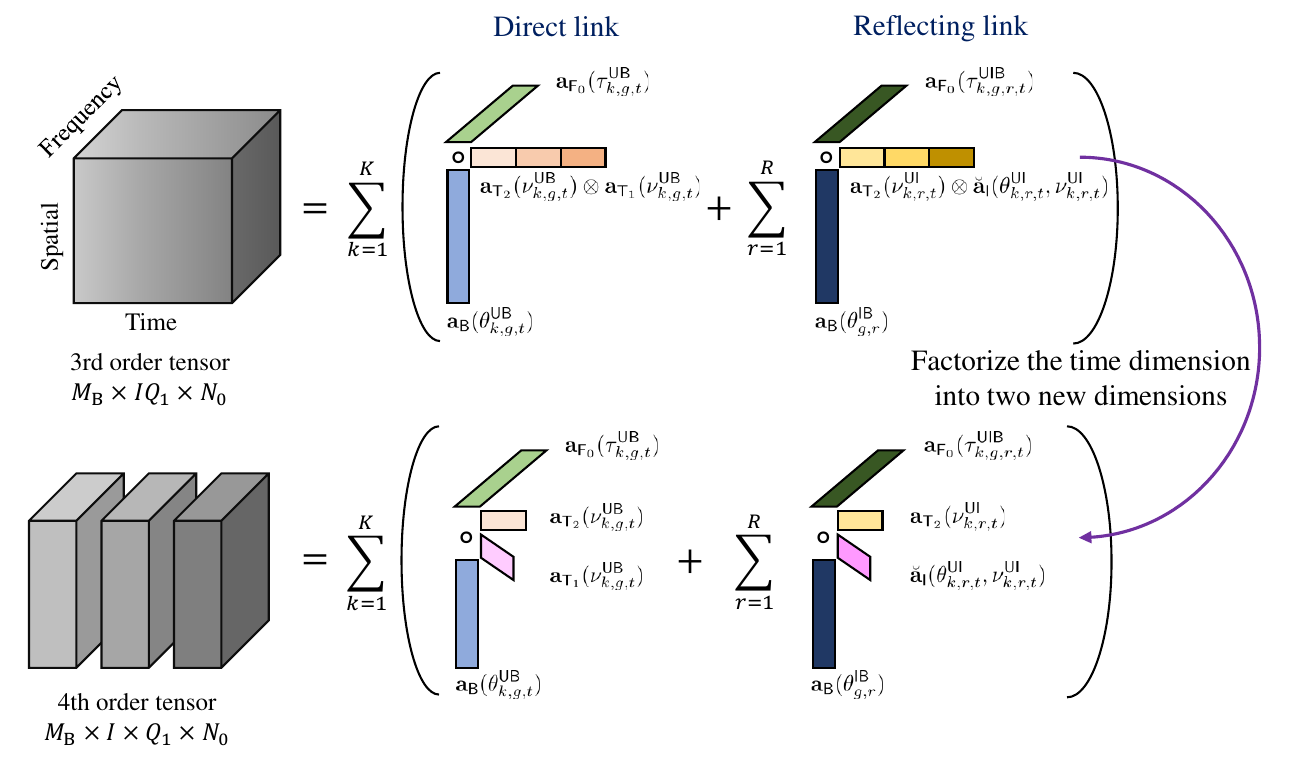}
  \vspace{-0.3cm}
  \caption{The tensor representation of the received signal.}\label{fig:sig_tensor}
  \vspace{-0.5cm}
\end{figure}


For the ISAC set, the received signals can be expressed as a 3rd order tensor $\pmb{\mathcal{Y}}_{g,t} \in \mathbb{C}^{M_{\mathsf{B}} \times Q_{\rm {I}} \times N_{\rm {I}}}$. 
Under the above design, we can further factorize the time dimension into two new dimensions, resulting in a 4th order tensor representation, as shown in Fig. \ref{fig:sig_tensor}. For convenience, we define the vector form of the received signal as
\begin{align}
    \mathbf{y}_{g,t} &= \text{vec}(\pmb{\mathcal{Y}}_{g,t}) \notag \\
    &= \left[(\mathbf{y}_{g,\varsigma_{1},t})^\mathsf{T},\dots,(\mathbf{y}_{g,\varsigma_{N_{\rm I}},t})^\mathsf{T}\right]^\mathsf{T}, ~\forall g,t, \label{equ:new_y}
\end{align}
where
\begin{align}
    \mathbf{y}_{g,\varsigma_{\breve{n}},t} = \bigg[&\Big(\mathbf{y}^{(\iota_{1,1})}_{g,\varsigma_{\breve{n}},t}\Big)^\mathsf{T}, \dots, \Big(\mathbf{y}^{(\iota_{1,Q_1})}_{g,\varsigma_{\breve{n}},t}\Big)^\mathsf{T},  \dots, \notag \\
     &\Big(\mathbf{y}^{(\iota_{I,1})}_{g,\varsigma_{\breve{n}},t}\Big)^\mathsf{T}, \dots, \Big(\mathbf{y}^{(\iota_{I,Q_1})}_{g,\varsigma_{\breve{n}},t}\Big)^\mathsf{T} \bigg]^\mathsf{T}, ~\forall g,\tilde{q},t.
\end{align}
Specifically, the signal vector \eqref{equ:new_y} can be re-expressed by
\begin{align}
    \mathbf{y}_{g,t} =&~ \sum_{k=1}^{K} \bigg( w^{\mathsf{UB}}_{k,g,t} \Big(\mathbf{a}_{\mathsf{F}}(\tau^{\mathsf{UB}}_{k,g,t}) \otimes \breve{\mathbf{a}}_{{\mathsf{T}}}(\nu^{\mathsf{UB}}_{k,g,t}) \otimes \mathbf{a}_{{\mathsf{T}}}(\nu^{\mathsf{UB}}_{k,g,t}) \notag \\
    &\otimes \mathbf{a}_{\mathsf{B}}(\theta^{\mathsf{UB}}_{k,g,t})\Big) + \sum_{r=1}^{R} w^{\mathsf{UI}}_{k,g,r,t} \Big(\mathbf{a}_{\mathsf{F}}(\tau^{\mathsf{UIB}}_{k,g,r,t}) \otimes \breve{\mathbf{a}}_{{\mathsf{T}}}(\nu^{\mathsf{UI}}_{k,r,t}) \notag \\
    & \otimes \breve{\mathbf{a}}_{{\mathsf{I}}}(\theta^{\mathsf{I}}_{r},\theta^{\mathsf{UI}}_{k,r,t}, \nu^{\mathsf{UI}}_{k,r,t}) \otimes \mathbf{a}_{\mathsf{B}}(\theta^{\mathsf{I}}_{g,r}) \Big) \bigg) + \mathbf{z}_{g,t} \notag \\
    =&~ \sum_{k=1}^{K} \Big( w^{\mathsf{UB}}_{k,g,t} \mathbf{a}^{\mathsf{UB}}_{k,g,t} + \sum_{r=1}^{R} w^{\mathsf{UI}}_{k,g,r,t} \mathbf{a}^{\mathsf{UI}}_{k,g,r,t} \Big) + \mathbf{z}_{g,t} \notag \\
    =&~ \mathbf{A}_{g,t}\mathbf{w}_{g,t} + \mathbf{z}_{g,t}, \label{equ:y_gt_vec}
\end{align}
where $w_{k,g,t}^{\mathsf{UB}} = \sqrt{P}\tilde{s}_{k,t} \alpha^{{\mathsf{UB}}}_{k,g,t} \beta^{\mathsf{UB}}_{k,g,t}$ and $w_{k,g,r,t}^{\mathsf{UI}} = \sqrt{P}\tilde{s}_{k,t} \alpha^{{\mathsf{IB}}}_{g,r} \alpha^{{\mathsf{UI}}}_{k,r,t} \beta^{\mathsf{IB}}_{g,r} \beta^{\mathsf{UI}}_{k,r,t}$ are effective signals along the user $k$-BS $g$ and user $k$-RIS $r$-BS $g$ links, respectively; $\tau^{\mathsf{UIB}}_{k,g,r,t} = \tau^{\mathsf{UI}}_{k,r,t} + \tau^{\mathsf{IB}}_{g,r}$ is the total delay of the user $k$-RIS $r$-BS $g$ link; 
$\mathbf{a}^{\mathsf{UB}}_{k,g,t} = \mathbf{a}_{\mathsf{F}_0}(\tau^{\mathsf{UB}}_{k,g,t}) \otimes \mathbf{a}_{{\mathsf{T}}_2}(\nu^{\mathsf{UB}}_{k,g,t}) \otimes \mathbf{a}_{{\mathsf{T}}_1}(\nu^{\mathsf{UB}}_{k,g,t}) \otimes \mathbf{a}_{\mathsf{S}}(\theta^{\mathsf{UB}}_{k,g,t})$ is the effective steering vector for the user $k$-BS $g$ channel and $\mathbf{a}^{\mathsf{UI}}_{k,g,r,t}$ is similarly defined; $\mathbf{A}_{g,t} = [\mathbf{A}_{1,g,t},\dots,\mathbf{A}_{K,g,t}]$ and $\mathbf{w}_{g,t} = [(\mathbf{w}_{1,g,t})^\mathsf{T},\dots,(\mathbf{w}_{K,g,t})^\mathsf{T}]^\mathsf{T}$ are the effective steering matrix and effective signal vector, respectively, with $\mathbf{A}_{k,g,t} = [\mathbf{a}^{\mathsf{UB}}_{k,g,t},\mathbf{a}^{\mathsf{UI}}_{k,g,1,t}\dots,\mathbf{a}^{\mathsf{UI}}_{k,g,R,t}]$ and $\mathbf{w}_{k,g,t} = [w^{\mathsf{UB}}_{k,g,t},w^{\mathsf{UI}}_{k,g,1,t}\dots,w^{\mathsf{UI}}_{k,g,R,t}]^\mathsf{T}$. Then, the aggregated signal matrix from all BSs can be expressed by $\mathbf{Y}_t = [\mathbf{y}_{1,t},\dots,\mathbf{y}_{G,t}]$.
In \eqref{equ:y_gt_vec}, $\breve{\mathbf{a}}_{\mathsf{I}}(\cdot)$, $\mathbf{a}_{\mathsf{F}}(\cdot)$, and $\mathbf{a}_{\mathsf{T}}(\cdot)$/$\breve{\mathbf{a}}_{\mathsf{T}}(\cdot)$ are the RIS-related, delay-related, and Doppler-related steering vectors defined by\footnote{When $Q_1$ is not very large, we observe that $\mathbf{a}_{\mathsf{T}_1}(\nu) \approx \mathbf{1}_{Q_1}$, which can be leveraged to simplify the signal model. From our simulation trials, we find that there has little influence on the joint estimation performance by adopting this approximation.}
\begin{align}
  \breve{\mathbf{a}}_{\mathsf{I}}(\varphi,\theta,\nu) &= (\pmb{\Psi}_{r,t}^\mathsf{T} (\mathbf{a}_{\mathsf{I}}(\varphi)\odot\mathbf{a}_{\mathsf{I}}(\theta))) \odot \mathbf{a}_{{\mathsf{T}}}(\nu), \notag \\
  &= (\pmb{\Psi}_{r,t}^\mathsf{T} \tilde{\mathbf{a}}_{\mathsf{I}}(\varphi, \theta))) \odot \mathbf{a}_{{\mathsf{T}}}(\nu), \label{equ:a_I_gen} \\
  \mathbf{a}_{\mathsf{F}}(\tau) &= \big[ 1, e^{-\jmath \breve{\zeta}_{\mathsf{F}} \tau}, \dots, e^{-\jmath \breve{\zeta}_{\mathsf{F}} (N_0 - 1) \tau} \big]^\mathsf{T}, \\
  \mathbf{a}_{{\mathsf{T}}}(\nu)  &= \big[ 1, e^{-\jmath \zeta_{{\mathsf{T}}} \nu},    \dots, e^{-\jmath \zeta_{{\mathsf{T}}} (Q_1 - 1) \nu} \big]^\mathsf{T}, \\
  \breve{\mathbf{a}}_{{\mathsf{T}}}(\nu)  &= \big[ 1, e^{-\jmath \breve{\zeta}_{{\mathsf{T}}} \nu},    \dots, e^{-\jmath \breve{\zeta}_{{\mathsf{T}}} (I - 1) \nu} \big]^\mathsf{T},
\end{align}
respectively, where $\pmb{\Psi}_r = [\tilde{\pmb{\phi}}_{r,t}^{(1)},\dots,\tilde{\pmb{\phi}}_{r,t}^{(Q_1)}], ~\forall r$, $ \tilde{\mathbf{a}}_{\mathsf{I}}(\varphi, \theta) =  \mathbf{a}_{\mathsf{I}}(\varphi_r)\odot\mathbf{a}_{\mathsf{I}}(\theta)$, $\breve{\zeta}_{\mathsf{F}} = \zeta_{\mathsf{F}}\Delta N$, and $\breve{\zeta}_{{\mathsf{T}}} = \zeta_{\mathsf{T}}\Delta Q$.

\subsection{Probabilistic Signal Representation}\label{subsec:prob_SR}

With the above, we establish the probabilistic signal model in the considered system, which is then applied to the Bayesian problem formulation.
We first apply the geometric constraints to build the statistical relationship between position-related parameters and user states. Using the user $r$-BS $g$ channel, $\forall r, g$, as an example, we have
\begin{align}
  p(\theta_{k,g,t}^{\mathsf{UB}}|\pmb{\psi}_{k,t}) &= \delta\left(\theta_{k,g,t}^{\mathsf{UB}} - \arccos( (\mathbf{e}^{\mathsf{B}}_{g})^\mathsf{T} \mathbf{e}^{\mathsf{UB}}_{k,g,t})\right), \label{equ:p_theta_state} \\
  p(\tau^{\mathsf{UB}}_{k,g,t}|\pmb{\psi}_{k,t}) &= \delta\left(\tau_{k,g,t}^{\mathsf{UB}} - c_0^{-1}d_{k,g,t}^{\mathsf{UB}}\right), \label{equ:p_tau_state} \\
  p(\nu^{\mathsf{UB}}_{k,g,t}|\pmb{\psi}_{k,t}) &= \delta\left(\nu^{\mathsf{UB}}_{k,g,t} - \lambda^{-1} (\dot{\mathbf{p}}_{k,t}^{\mathsf{U}})^\mathsf{T} \mathbf{e}_{k,g,t}^{\mathsf{UB}}, \label{equ:p_nu_state}  \right),
\end{align}
where $d_{k,g,t}^{\mathsf{UB}} = ||\mathbf{p}_{k,t}^{\mathsf{U}} - \mathbf{p}^{\mathsf{B}}_{g}||_2, ~\forall k,t$, is the distance between user $k$ and the BS $g$ at the time slot $t$, $\mathbf{e}^{\mathsf{UB}}_{k,g,t} = (d_{k,g,t}^{\mathsf{UB}})^{-1}(\mathbf{p}_{k,t}^{\mathsf{U}} - \mathbf{p}^{\mathsf{B}}_{g})$ is the directional vector from the BS $g$ to the user $k$ at the time slot $t$, and $\mathbf{e}^{\mathsf{B}}_{g} \in \mathbb{R}^{2 \times 1}$ is the array directional vector of the BS $g$.
The statistical relationships for user $k$-RIS $r$ and RIS $r$-BS $g$ channels are similar to \eqref{equ:p_theta_state} -- \eqref{equ:p_nu_state}.
Note that the Doppler frequency for each RIS-BS channel is equal to zero as shown in \eqref{equ:h_IB}, due to the static nature of RISs and BSs.
Based on the mobility model \eqref{equ:discrete_mm}, the state of each user $k$ is modeled to evolve according to a first-order Markov process as
\begin{align}\label{equ:mm_markov}
  p(\pmb{\psi}_{k,t}|\pmb{\psi}_{k,t-1}) &= \mathcal{N}(\pmb{\psi}_{k,t};\mathbf{F}_0\pmb{\psi}_{k,t-1},\mathbf{Q}_{k,t-1}), ~\forall k,t.
\end{align}

We denote $\pmb{\vartheta}_{t} = [(\pmb{\vartheta}_{1,t})^\mathsf{T},\dots,(\pmb{\vartheta}_{K, t})]^\mathsf{T}$ as the position-related parameter vector for all users in the time slot $t$. Therein, the position-related parameter vector of each user $k$ is defined by 
\begin{align}
  \pmb{\vartheta}_{k,t} =&~ [(\pmb{\vartheta}^{\mathsf{UB}}_{k,1,t})^\mathsf{T},\dots,(\pmb{\vartheta}^{\mathsf{UB}}_{k,G,t})^\mathsf{T}, \notag \\
  &~~(\pmb{\vartheta}^{\mathsf{UI}}_{k,1,t})^\mathsf{T},\dots,(\pmb{\vartheta}^{\mathsf{UI}}_{k,R,t})^\mathsf{T}]^\mathsf{T}, ~\forall k,t,
\end{align}
where $\pmb{\vartheta}^{\mathsf{c}}_{k,g,t} = [\theta^{\mathsf{c}}_{k,g,t},\tau^{\mathsf{c}}_{k,g,t},\nu^{\mathsf{c}}_{k,g,t}]^\mathsf{T}, ~\forall \mathsf{c} \in \{\mathsf{UB}, \mathsf{UI}\}$ 
are the position-related parameter vector for each link. We also denote the link indicator vector as $\pmb{\alpha}_{t} = [\pmb{\alpha}_{1,t}^\mathsf{T},\dots,\pmb{\alpha}_{G,t}^\mathsf{T}]^\mathsf{T}$ with $\pmb{\alpha}_{g,t} = [\pmb{\alpha}_{1,g,t}^\mathsf{T},\dots,\pmb{\alpha}_{K,g,t}^\mathsf{T}]^\mathsf{T}$ and $\pmb{\alpha}_{k,g,t} = [\alpha^{\mathsf{UB}}_{k,g,t},\alpha^{\mathsf{UI}}_{k,g,1,t},\dots,\alpha^{\mathsf{UI}}_{k,g,R,t}]^\mathsf{T}$.
By following \eqref{equ:y_gt_vec}, the conditional probability for the received signal at the BS $g$ in the ISAC sub-block of each time slot $t$ over the position-related parameters of all users can be given by
\begin{align}\label{equ:cond_p_y_state}
  p(\mathbf{y}_{g,t}|\pmb{\vartheta}_{t}, \tilde{\mathbf{s}}_{t},\pmb{\alpha}_{g,t}) = \mathcal{CN}(\mathbf{y}_{g,t};\mathbf{A}_{g,t}\mathbf{w}_{g,t},\sigma^2_z \mathbf{I}).
\end{align}
The probability density function (PDF) of the transmit signal is also denoted as $\mathcal{P}_0(\tilde{s}_{k,t}), ~\forall k,t$. 
To perform online estimation, we can derive the joint probability of the received signals, transmit signals, states, and position-related parameters from the first time slot to the $t$-th time slot as
\begin{align}\label{equ:joint_prob}
    &p(\{\mathbf{Y}_{v},\pmb{\vartheta}_{v}, \pmb{\psi}_{v}, \tilde{\mathbf{s}}_{v}, \pmb{\alpha}_{v} \}_{v=1}^{t}) = \prod_{v=1}^{t} p(\mathbf{y}_{g,v}|\pmb{\vartheta}_{v}, \tilde{\mathbf{s}}_{v},\pmb{\alpha}_{g,v}) \notag \\
    & \times \bigg( \prod_{k=1}^{K} p(\pmb{\vartheta}_{k,v}| \pmb{\psi}_{k,v})  p(\pmb{\psi}_{k,v}|\pmb{\psi}_{k,v-1}) \mathcal{P}_0(\tilde{s}_{k,v}) \bigg),
\end{align}
where $p(\pmb{\psi}_{k,1}|\pmb{\psi}_{k,0}) = p(\pmb{\psi}_{k,1})$ denotes the prior information of the user states at the first time slot derived from the initial estimation of user states. 

\subsection{Bayesian Problem Formulation}

Under the probabilistic model, we can follow the Bayes' rule to derive the posterior probabilities of the state and transmit signal of each user $k$ in the time slot $t$ as 
\begin{small}
\begin{align}
    p(\pmb{\psi}_{k,t}|\{\mathbf{Y}_v\}_{v = 1}^{t}) &= \int_{\setminus \pmb{\psi}_{k,t}} \frac{p(\{\mathbf{Y}_{v},\pmb{\vartheta}_{v}, \pmb{\psi}_{v}, \tilde{\mathbf{s}}_{v}, \pmb{\alpha}_{v} \}_{v = 1}^{t})}{p(\{\mathbf{Y}_{v}\}_{v = 1}^{t})}, \label{equ:postprob_state}  \\
    p(\tilde{s}_{k,t}|\{\mathbf{Y}_v\}_{v = 1}^{t}) &= \int_{\setminus \tilde{s}_{k,t}}\frac{ p(\{\mathbf{Y}_{v},\pmb{\vartheta}_{v}, \pmb{\psi}_{v}, \tilde{\mathbf{s}}_{v}, \pmb{\alpha}_{v} \}_{v = 1}^{t})}{p(\{\mathbf{Y}_{v}\}_{v = 1}^{t})}. \label{equ:postprob_signal}
\end{align}
\end{small}After that, we can apply the criterion of MAP or MMSE to perform the online estimation at the time slot $t$. Meanwhile, the position-related parameters and link indicators can also be derived simultaneously. However, it is intractable to derive the explicit expression of \eqref{equ:postprob_state} -- \eqref{equ:postprob_signal} due to high-dimensional integrals. Though the popular KF-based methods \cite{Einicke_1999_TSP} can approximately derive a solution, their performance suffers severe deterioration in the case with limited tracking resources. As such, this work resorts to leveraging MP and VMP techniques under the Bayesian approximate inference framework to design an effective and efficient estimation algorithm.

\section{Joint Multiuser Tracking and Signal Detection Algorithm}

In this section, we propose a computationally efficient HVMP algorithm for the joint multi-user tracking and signal detection problem based on the built probabilistic model.

\begin{figure}[t]
  \centering
  \includegraphics[width=.42\textwidth]{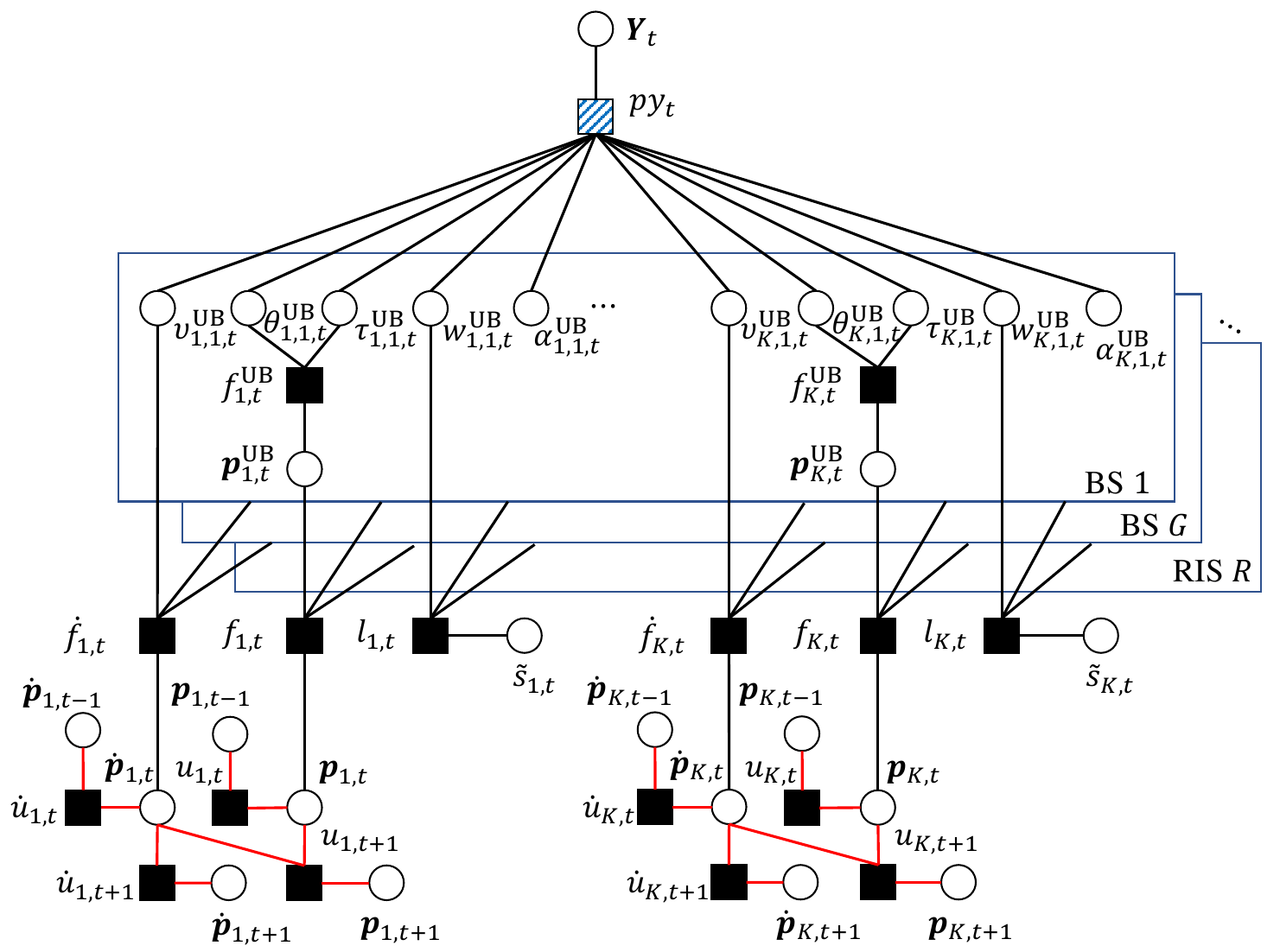}
  \vspace{-0.3cm}
  \caption{Factor graph of the probabilistic model.}\label{Fig:factor_graph}
  \vspace{-0.5cm}
\end{figure}

\subsection{Graphical Representation}
Before the introduction of the proposed algorithm, we first establish the graphical representation of factor graph based on the joint probability \eqref{equ:joint_prob} as shown in Fig. \ref{Fig:factor_graph}. Therein, squares and white circles denote the factor nodes and variable nodes, respectively. Moreover, we propose to perform bidirectional message passing over the (black) edges for variable update in each time slot, and perform only forward message passing over the (red) edges for state update, due to the online estimation. Note that variational message passing is performed along edges connected with the blue-veined square and standard message passing is realized along the other edges. In the following, we introduce the details of the proposed algorithm. 

\subsection{Hybrid Variational Message Passing}

In general, the proposed HVMP algorithm integrates both the variational message passing and standard message passing to iteratively update the extrinsic message on each parameter, which are utilized to derive the posterior probability of user states and transmit signals after convergence. Specifically, the VMP is applied to estimate position-related parameters from the received signals, while the MP is utilized to refine the estimation of states and transmit signals. By employing such a hybrid of VMP and MP rules, the proposed algorithm can update all the extrinsic messages with tractable expressions to realize computationally efficient online estimation for user states and transmit signals.

We first introduce the bidirectional message passing equations sequentially and then give the forward message passing equations for online estimation. By leveraging the variational techniques in the message passing procedure, we consider to model distributions with both of the position and velocity by Gaussian distributions, while the von Misés (VM) distribution \cite{Badiu_2017_TSP,Gao_2025_TWC} is used for modeling position-related parameters.

\subsubsection{Messages between $\dot{\mathbf{p}}_{k,t}$ and $\dot{f}_{k,t}$}
The bidirectional messages in the $j$-th iteration can be given by
\begin{small}
\begin{align}
    \varepsilon_{\dot{\mathbf{p}}_{k,t} \rightarrow \dot{f}_{k,t}}(j) &= \mathcal{N}(\dot{\mathbf{p}}_{k,t}; \dot{\overrightarrow{\mathbf{p}}}_{k,t}, \dot{\overrightarrow{\mathbf{Q}}}_{k,t}), \label{equ:m_dp_to_df} \\
    \varepsilon_{\dot{\mathbf{p}}_{k,t} \leftarrow \dot{f}_{k,t}}(j)   &\varpropto \int p(\{\nu^{\mathsf{UB}}_{k,g,t}\},\{\nu^{\mathsf{UI}}_{k,r,t}\}|\dot{\mathbf{p}}_{k,t}) \notag \\
    & ~~~ \times \prod_{g=1}^{G} \varepsilon_{\dot{f}_{k,t} \leftarrow \nu^{\mathsf{UB}}_{k,g,t}}(j) \prod_{r=1}^{R} \varepsilon_{\dot{f}_{k,t} \leftarrow \nu^{\mathsf{UI}}_{k,r,t}}(j) \notag \\
    & = \mathcal{N}(\dot{\mathbf{p}}_{k,t}; \dot{\overleftarrow{\mathbf{p}}}_{k,t}(j), \dot{\overleftarrow{\mathbf{Q}}}_{k,t}(j)),  \label{equ:m_df_to_dp}
\end{align}
\end{small}where $\dot{\overrightarrow{\mathbf{p}}}_{k,t}$ and $\dot{\overrightarrow{\mathbf{Q}}}_{k,t}$ are the mean and covariance of the velocity from the online estimation in the last time slot $t-1$, $\dot{\overleftarrow{\mathbf{p}}}_{k,t}(j)$ and $\dot{\overleftarrow{\mathbf{Q}}}_{k,t}(j)$ are the updated ones from the Doppler frequency estimation in the $j$-th iteration of the HVMP algorithm. The explicit expression of $\dot{\overleftarrow{\mathbf{p}}}_{k,t}(j)$ and $\dot{\overleftarrow{\mathbf{Q}}}_{k,t}(j)$ is hard to be derived due to the complicate relationship between the Doppler frequencies and the velocity, and we resort to the Gaussian-Newton method and Taylor series expansion to obtain $\dot{\overleftarrow{\mathbf{p}}}_{k,t}(j)$ and $\dot{\overleftarrow{\mathbf{Q}}}_{k,t}(j)$. The detailed derivation is provided in Appendix \ref{sec:message_dp}.

\subsubsection{Messages between $\mathbf{p}_{k,t}$ and $f_{k,t}$}
The bidirectional messages in the $j$-th iteration are derived by
\begin{small}
\begin{align}
    \varepsilon_{\mathbf{p}_{k,t} \rightarrow f_{k,t}}(j) =&~ \mathcal{N}(\mathbf{p}_{k,t}; \overrightarrow{\mathbf{p}}_{k,t}, \overrightarrow{\mathbf{Q}}_{k,t}), ~\forall j, \label{equ:m_p_to_f} \\
    \varepsilon_{\mathbf{p}_{k,t} \leftarrow f_{k,t}}(j)   \varpropto&~ \int p(\{\mathbf{p}^{\mathsf{UB}}_{k,g,t}\},\{\mathbf{p}^{\mathsf{UI}}_{k,r,t}\}|\mathbf{p}_{k,t}) \sum_{g=1}^{G} \varepsilon_{f_{k,t} \leftarrow \mathbf{p}^{\mathsf{UB}}_{k,g,t}}(j) \notag \\
    &\times \prod_{r=1}^{R} \varepsilon_{f_{k,t} \leftarrow \mathbf{p}^{\mathsf{UI}}_{k,r,t}}(j) \notag \\
    =&~ \mathcal{N}(\mathbf{p}_{k,t}; \overleftarrow{\mathbf{p}}_{k,t}(j), \overleftarrow{\mathbf{Q}}_{k,t}(j)), \label{equ:mp_fusion_p}
\end{align}
\end{small}
where 
\begin{small}
\begin{align}
  \overleftarrow{\mathbf{p}}_{k,t}(j) =&~ \Big(\overleftarrow{\mathbf{Q}}_{k,t}(j) \Big)^{-1} \bigg[ \sum_{g=1}^{G} \Big(\overleftarrow{\mathbf{Q}}^{{\mathsf{UB}}}_{k,g,t}(j) \Big)^{-1} \overleftarrow{\mathbf{p}}^{{\mathsf{UB}}}_{k,g,t}(j)  \\
  &+ \sum_{r=1}^{R} \Big(\overleftarrow{\mathbf{Q}}^{{\mathsf{UI}}}_{k,r,t}(j) \Big)^{-1} \overleftarrow{\mathbf{p}}^{{\mathsf{UI}}}_{k,r,t}(j) \bigg] , \notag \\
  \overleftarrow{\mathbf{Q}}_{k,t}(j) =&~ \bigg( \sum_{g=1}^{G} \Big(\overleftarrow{\mathbf{Q}}^{{\mathsf{UB}}}_{k,g,t}(j) \Big)^{-1} + \sum_{r=1}^{R}  \Big(\overleftarrow{\mathbf{Q}}^{{\mathsf{UI}}}_{k,r,t}(j) \Big)^{-1} \bigg)^{-1} . 
\end{align}
\end{small}In the above messages, $\overrightarrow{\mathbf{p}}_{k,t}$ ($\overrightarrow{\mathbf{Q}}_{k,t}$) is the mean (covariance) of the position from the online estimation in the last time slot $t-1$ and the explicit expression of the message  $\varepsilon_{f_{k,t} \leftarrow \mathbf{p}^{\mathsf{UB}}_{k,g,t}}(j)$ ($\varepsilon_{f_{k,t} \leftarrow \mathbf{p}^{\mathsf{UI}}_{k,r,t}}(j)$) is given in \eqref{equ:m_pUB_to_f}.

\subsubsection{Messages between $\tilde{s}_{k,t}$ and $l_{k,t}$}
The message from $\tilde{s}_{k,t}$ to $l_{k,t}$ is obtained based on the prior distribution of the transmit signal, i.e., $\varepsilon_{\tilde{s}_{k,t} \rightarrow l_{k,t}}(j) = \mathcal{P}_0(\tilde{s}_{k,t})$, and the message from $l_{k,t}$ to ${\tilde{s}}_{k,t}$ is derived by
\begin{small}
\begin{align}
  \varepsilon_{\tilde{s}_{k,t} \leftarrow l_{k,t}}(j) \varpropto&~ \int p(\{w^{\mathsf{UB}}_{k,g,t}\}, \{w^{\mathsf{UI}}_{k,r,t}\}| \tilde{s}_{k,t}) \prod_{g=1}^{G}  \varepsilon_{l_{k,t}  \leftarrow w^{\mathsf{UB}}_{k,g,t}}(j) \notag \\
  & \times \prod_{r=1}^{R} \varepsilon_{l_{k,t} \leftarrow w^{\mathsf{UI}}_{k,r,t}}(j) \notag \\
  =&~ \mathcal{CN}(\tilde{s}_{k,t}; \overleftarrow{s}_{k,t}(j),\overleftarrow{\rho}_{k,t}(j)),  ~\forall j, \label{equ:m_s_bto_l}
\end{align}
\end{small}where 
\begin{small}
\begin{align}
  \overleftarrow{s}_{k,t}(j) =&~ (\overleftarrow{\rho}_{k,t}(j))^{-1} \bigg( \sum_{g=1}^{G} (\overleftarrow{\varpi}^{{\mathsf{UB}}}_{k,g,t}(j))^{-1}\overleftarrow{w}^{{\mathsf{UB}}}_{k,g,t}(j)  \\
  & + \sum_{r=1}^{R} (\overleftarrow{\varpi}^{{\mathsf{UI}}}_{k,r,t}(j))^{-1}\overleftarrow{w}^{{\mathsf{UI}}}_{k,r,t}(j)  \bigg), \notag \\
  \overleftarrow{\rho}_{k,t} =&~ \bigg((\overleftarrow{\varpi}^{{\mathsf{UB}}}_{k,g,t}(j))^{-1} +  \sum_{r=1}^{R} (\overleftarrow{\varpi}^{{\mathsf{UI}}}_{k,r,t}(j) )^{-1} \bigg)^{-1},
\end{align}
\end{small}with $\varepsilon_{l_{k,t}  \leftarrow w^{\mathsf{UB}}_{k,g,t}}(j) = \mathcal{CN}(w^{\mathsf{UB}}_{k,g,t}; \overleftarrow{w}^{\mathsf{UB}}_{k,g,t}(j),\overleftarrow{\varpi}^{\mathsf{UB}}_{k,g,t}(j))$ and $\varepsilon_{l_{k,t}  \leftarrow w^{\mathsf{UI}}_{k,r,t}}(j) = \mathcal{CN}(w^{\mathsf{UI}}_{k,r,t}; \overleftarrow{w}^{\mathsf{UI}}_{k,r,t}(j),\overleftarrow{\varpi}^{\mathsf{UI}}_{k,r,t}(j))$.

\subsubsection{Messages between $\dot{f}_{k,t}$ and $\nu^{\mathsf{UB}}_{k,g,t}$ ($\nu^{\mathsf{UI}}_{k,r,t}$)}
The bidirectional message passing equations can be written as
\begin{align}
  &\varepsilon_{\dot{f}^{\mathsf{UB}}_{k,t} \rightarrow \nu^{\mathsf{UB}}_{k,t}}(j) \varpropto \int p(\{\nu^{\mathsf{UB}}_{k,g,t}\},\{\nu^{\mathsf{UI}}_{k,r,t}\}|\dot{\mathbf{p}}_{k,t}) \notag \\
  & \cdot \prod_{g' \ne g} \varepsilon_{g_{k,t} \leftarrow \nu^{\mathsf{UB}}_{k,g',t}}(j) \prod_{r=1}^{R} \varepsilon_{g_{k,t} \leftarrow \nu^{\mathsf{UI}}_{k,r,t}}(j) \quad\quad \notag \\
  &= \mathcal{VM}(-\zeta_{\mathsf{T}} \nu^{\mathsf{UB}}_{k,t}; \overrightarrow{\mu}^{\mathsf{UB}}_{k,g,t}(\nu, j), \overrightarrow{\kappa}^{\mathsf{UB}}_{k,g,t}(\nu, j)), \label{equ:m_df_to_nu} \\
  &\varepsilon_{\dot{f}^{\mathsf{UB}}_{k,t} \leftarrow \nu^{\mathsf{UB}}_{k,g,t}}(j) = \varepsilon_{\nu^{\mathsf{UB}}_{k,g,t} \leftarrow py_{t}}(j) \notag \\
  &= \mathcal{VM}(-\zeta_{\mathsf{T}} \nu^{\mathsf{UB}}_{k,t}; \overleftarrow{\mu}^{\mathsf{UB}}_{k,g,t}(\nu, j), \overleftarrow{\kappa}^{\mathsf{UB}}_{k,g,t}(\nu, j)), \label{equ:m_nu_to_df}
\end{align}
where $\overrightarrow{\mu}^{\mathsf{UB}}_{k,g,t}(\nu, j)$ and $\overrightarrow{\kappa}^{\mathsf{UB}}_{k,g,t}(\nu, j))$ are also derived by the Gauss-Newton method as detailed in Appendix \ref{sec:message_dp}; $\varepsilon_{\dot{f}^{\mathsf{UB}}_{k,t} \leftarrow \nu^{\mathsf{UB}}_{k,g,t}}(j)$ ($\varepsilon_{\dot{f}^{\mathsf{UI}}_{k,t} \leftarrow \nu^{\mathsf{UI}}_{k,r,t}}(j)$) is derived from the variational message passing. The messages between $\dot{f}_{k,t}$ and $\nu^{\mathsf{UI}}_{k,r,t}$ can be similarly obtained to \eqref{equ:m_df_to_nu} and \eqref{equ:m_nu_to_df}.

\subsubsection{Messages between $f_{k,t}$ and $\mathbf{p}^{\mathsf{UB}}_{k,g,t}$ ($\mathbf{p}^{\mathsf{UI}}_{k,r,t}$)}
Due to the Gaussian message passing along the edges connected with node $f_{k,t}$, we can simply derive the message $\varepsilon_{f_{k,t} \rightarrow \mathbf{p}^{\mathsf{UB}}_{k,g,t}}(j)$ ($\varepsilon_{f_{k,t} \rightarrow \mathbf{p}^{\mathsf{UI}}_{k,r,t}}(j)$) similar to \eqref{equ:mp_fusion_p} as 
\begin{small}
\begin{align}\label{equ:m_pUB_bto_f}
  \varepsilon_{f_{k,t} \rightarrow \mathbf{p}^{\mathsf{UB}}_{k,g,t}}(j) \propto&~ \int p(\{\mathbf{p}^{\mathsf{UB}}_{k,g,t}\},\{\mathbf{p}^{\mathsf{UI}}_{k,r,t}\}|\mathbf{p}_{k,t}) \varepsilon_{\mathbf{p}_{k,t} \rightarrow f_{k,t}}(j) \notag \\ 
  &\times \prod_{g' \ne g} \varepsilon_{f_{k,t} \leftarrow \mathbf{p}^{\mathsf{UB}}_{k,g',t}}(j) \prod_{r=1}^{R} \varepsilon_{f_{k,t} \leftarrow \mathbf{p}^{\mathsf{UI}}_{k,r,t}}(j) \notag \\
  =&~ \mathcal{N}(\mathbf{p}^{\mathsf{UB}}_{k,g,t}; \overrightarrow{\mathbf{p}}^{\mathsf{UB}}_{k,g,t}(j), \overrightarrow{\mathbf{Q}}^{\mathsf{UB}}_{k,g,t}(j)),
\end{align}
\end{small}and the message $\varepsilon_{f_{k,t} \leftarrow \mathbf{p}^{\mathsf{UB}}_{k,g,t}}(j)$ ($\varepsilon_{f_{k,t} \leftarrow \mathbf{p}^{\mathsf{UI}}_{k,r,t}}(j)$) can be obtained as
\begin{align}\label{equ:m_pUB_to_f}
  \varepsilon_{f_{k,t} \leftarrow \mathbf{p}^{\mathsf{UB}}_{k,g,t}}(j) =&~ \int p(\theta^{\mathsf{UB}}_{k,g,t},\tau^{\mathsf{UB}}_{k,g,t}|\mathbf{p}^{\mathsf{UB}}_{k,g,t}) \notag \\
  &\times \varepsilon_{f^{\mathsf{UB}}_{k,g,t} \leftarrow \theta^{\mathsf{UB}}_{k,g,t}}(j) \varepsilon_{f^{\mathsf{UB}}_{k,g,t} \leftarrow \tau^{\mathsf{UB}}_{k,g,t}}(j)  \notag \\
  =&~ \mathcal{N}(\mathbf{p}^{\mathsf{UB}}_{k,g,t}; \overleftarrow{\mathbf{p}}^{{\mathsf{UB}}}_{k,g,t}(j),\overleftarrow{\mathbf{Q}}^{{\mathsf{UB}}}_{k,g,t}(j)),
\end{align}
where the mean $\overleftarrow{\mathbf{p}}^{{\mathsf{UB}}}_{k,g,t}(j)$ and covariance $\overleftarrow{\mathbf{Q}}^{{\mathsf{UB}}}_{k,g,t}(j)$ are derived based on second-order Taylor series expansion and Gaussian approximation based on the updated messages on $\theta^{\mathsf{UB}}_{k,g,t}$ and $\tau^{\mathsf{UB}}_{k,g,t}$ in the part of variational message passing. The detailed derivation of \eqref{equ:m_pUB_to_f} is omitted due to page limits. 
Similarly, the messages between $f_{k,t}$ and $\mathbf{p}^{\mathsf{UI}}_{k,r,t}$ can be obtained by following the above calculations.

\subsubsection{Messages between $l_{k,t}$ and $w^{\mathsf{UB}}_{k,g,t}$ ($w^{\mathsf{UI}}_{k,g,r,t}$)}
The bidirectional messages can be calculated by
\begin{small}
\begin{align}
  \varepsilon_{l_{k,t} \rightarrow w^{\mathsf{UB}}_{k,g,t}}(j) &\varpropto \int p(\{w^{\mathsf{UB}}_{k,g,t}\}, \{w^{\mathsf{UI}}_{k,g,r,t}\}| \tilde{s}_{k,t}) \notag \\
  & \quad \times \prod_{g' \ne g} \varepsilon_{l_{k,t} \leftarrow w^{\mathsf{UB}}_{k,g',t}}(j) \prod_{r=1}^{R} \varepsilon_{l_{k,t} \leftarrow w^{\mathsf{UI}}_{k,g,r,t}}(j), \label{equ:l_r_w} \\
  \varepsilon_{l_{k,t} \leftarrow w^{\mathsf{UB}}_{k,g,t}}(j)   &= \mathcal{CN}(w^{\mathsf{UB}}_{k,g,t}; \overleftarrow{w}^{{\mathsf{UB}}}_{k,g,t}(j) ,  \overleftarrow{\varpi}^{{\mathsf{UB}}}_{k,g,t}(j)),  \label{equ:l_l_w}
\end{align}
\end{small}where $\overleftarrow{w}^{{\mathsf{UB}},j}_{k,t}$ and $\overleftarrow{\varpi}^{{\mathsf{UB}},j}_{k,t}$ are derived from the variational message passing. Messages of the variable $w^{\mathsf{UI}}_{k,r,t}$ is similarly derived by following \eqref{equ:l_r_w} and \eqref{equ:l_l_w}.

\subsubsection{Messages between $f^{\mathsf{UB}}_{k,g,t}$ and $\theta^{\mathsf{UB}}_{k,g,t}/\tau^{\mathsf{UB}}_{k,g,t}$ (between $f^{\mathsf{UI}}_{k,r,t}$ and $\theta^{\mathsf{UI}}_{k,r,t}/\tau^{\mathsf{UI}}_{k,r,t}$)}
By following \eqref{equ:m_df_to_nu}, we can also obtain the messages $\varepsilon_{f^{\mathsf{UB}}_{k,g,t} \rightarrow \theta^{\mathsf{UB}}_{k,g,t}}(j)$ and $\varepsilon_{f^{\mathsf{UB}}_{k,g,t} \rightarrow \tau^{\mathsf{UB}}_{k,g,t}}(j)$ in the form of VM distributions. Alternatively, the messages $\varepsilon_{f^{\mathsf{UB}}_{k,t} \leftarrow \theta^{\mathsf{UB}}_{k,t}}(j) = \varepsilon_{\theta^{\mathsf{UB}}_{k,g,t} \leftarrow py_{t}}(j)$ and $\varepsilon_{f^{\mathsf{UB}}_{k,t} \leftarrow \tau^{\mathsf{UB}}_{k,t}}(j) = \varepsilon_{\tau^{\mathsf{UB}}_{k,t} \leftarrow py_{t}}(j)$ are obtained from the converged results for the variational message passing in the $i$-th outer iteration of HVMP. The message passing for user-RIS paths  can be similarly obtained.

\subsubsection{Messages between $py_{t}$ and $\pmb{\vartheta}^{\mathsf{UB}}_{k,g,t}$ $(\pmb{\vartheta}^{\mathsf{UI}}_{k,r,t})$ via variational message passing}\label{sec:VMP}
From the signal model \eqref{equ:y_gt_vec}, this part aims to solve the multidimensional line spectra estimation problem \cite{Zhang_2020_SPL} to derive the extrinsic messages over the variable vector $\pmb{\vartheta}_{t}$. Under the Bayesian inference framework, the standard sum-product algorithm is intractable to be employed in this part, since all the variables $\pmb{\vartheta}_t$ are coupled with complicated and non-linear dependencies in the received signals $\{\mathbf{y}_{g,t}\}$. Motivated by the variational line spectra inference methods in \cite{Badiu_2017_TSP,Zhang_2020_SPL,Teng_2023_JSAC}, the VMP is employed in this part to realize efficient derivation of the desired extrinsic messages. However, the methods in \cite{Badiu_2017_TSP,Zhang_2020_SPL} cannot be directly applied due to the beamspace signal structure in \eqref{equ:y_gt_vec}, where the received signal at the RIS cannot be fully observed and is projected into a low-dimensional subspace $\{\pmb{\Psi}^\mathsf{T}_{r}\}$, as shown in \eqref{equ:a_I_gen}. On the other hand, the variational method in \cite{Teng_2023_JSAC} concentrates on the single-BS case and cannot fully exploit the dependencies of each element in $\pmb{\vartheta}^{\mathsf{UB}}_{k,g,t}$ ($\pmb{\vartheta}^{\mathsf{UI}}_{k,r,t}$) by optimizing each element individually via coordinate descent. To address the above challenges, we propose to generalize the variational inference-based method to deal with the beamspace multidimensional line spectra estimation problem.

The objective of VMP is to approximate the PDF of the posterior probability over these variables by a surrogate PDF with the minimal Kullback-Leibler (KL) divergence\footnote{The KL divergence of $p_1(x)$ to $p_2(x)$ (within the same feasible set $\mathcal{X}$) is defined as $\mathcal{D}_{\rm KL}(p_1||p_2) = \int_{\mathcal{X}} p_1(x) \ln \frac{p_1(x)}{p_2(x)} dx$.}. 
Due to the fact that
\begin{align}
  \ln p_{\pmb{Y}}(\mathbf{Y}_t) &= \mathcal{D}_{\rm KL}(\chi_{\pmb{\vartheta},\pmb{W},\pmb{\alpha}|\pmb{Y}}||p_{\pmb{\vartheta},\pmb{W},\pmb{\alpha}|\pmb{Y}}) + \mathcal{L}(\chi_{\pmb{\vartheta},\pmb{W},\pmb{\alpha}|\pmb{Y}}), \notag \\
  \mathcal{L}(\chi_{\pmb{\vartheta},\pmb{W},\pmb{\alpha}|\pmb{Y}}) &= \mathbb{E}_{\chi_{\pmb{\vartheta},\pmb{W},\pmb{\alpha}|\pmb{Y}}}\left[ \ln \frac{p_{\pmb{\vartheta},\pmb{W},\pmb{\alpha},\pmb{Y}}(\pmb{\vartheta}_t,\mathbf{W}_t,\pmb{\alpha}_t,\mathbf{Y}_t)}{ \chi_{\pmb{\vartheta},\pmb{W},\pmb{\alpha}|\pmb{Y}}(\pmb{\vartheta}_t,\mathbf{W}_t,\pmb{\alpha}_t|\mathbf{Y}_t)} \right], \notag
\end{align}
we can turn to maximize the evidence lower bound (ELBO) $\mathcal{L}(\chi_{\pmb{\vartheta},\pmb{W},\pmb{\alpha}|\pmb{Y}})$ given that $p_{\pmb{Y}}(\mathbf{Y}_t)$ is constant for the received signal $\mathbf{Y}_t$.
Note that the variables $\{w^{\mathsf{UB}}_{k,g,t}\}$ and $\{\pmb{\vartheta}^{\mathsf{UB}}_{k,g,t}\}$ ($\{w^{\mathsf{UI}}_{k,g,r,t}\}$, $\{\pmb{\vartheta}^{\mathsf{UI}}_{k,r,t}\}$) are all considered to be independent in the VMP part.
Therefore, the surrogate PDF can be factorized into the product of a set of independent factors as 
\begin{align}\label{equ:sur_PDF_fac}
 & \chi_{\pmb{\vartheta},\pmb{W},\pmb{\alpha}|\pmb{Y}} = \prod_{g=1}^{G} \chi_{\pmb{W}|\pmb{Y}}(\mathbf{w}_{g,t}|\mathbf{Y}_t) \chi_{\pmb{\alpha}|\pmb{Y}}(\pmb{\alpha}_{g,t}|\mathbf{Y}_t) \notag \\
  & \times  \prod_{k=1}^{K} \left(\prod_{g=1}^{G} (\chi_{\pmb{\vartheta}|\pmb{Y}}(\pmb{\vartheta}^{\mathsf{UB}}_{k,g,t}|\mathbf{Y}_t) \prod_{r=1}^{R}\chi_{\pmb{\vartheta}|\pmb{Y}}(\pmb{\vartheta}^{\mathsf{UI}}_{k,r,t}|\mathbf{Y}_t))\right), 
\end{align}
where $\chi_{\pmb{\vartheta}|\pmb{Y}}(\pmb{\vartheta}_{k,t}|\mathbf{Y}_t) = \prod_{i=1}^{3} \mathcal{VM}([\pmb{\vartheta}_{k,t}]_i;[\hat{\pmb{\vartheta}}_{k,t}]_i,[\hat{\pmb{\kappa}}_{k,t}]_i)$, $\chi_{\pmb{W}|\pmb{Y}}(\mathbf{w}_{g,t}|\mathbf{Y}_t)$ is modeled by the Gaussian distribution, and $\chi_{\pmb{\alpha}|\pmb{Y}}(\pmb{\alpha}_{g,t}|\mathbf{Y}_t) = \prod_{i} \delta([\pmb{\alpha}_{g,t}]_{i} - [\hat{\pmb{\alpha}}_{g,t}]_{i})$. 
To maximize the ELBO, we resort to the efficient block coordinate descent (BCD) method to optimize each factor in \eqref{equ:sur_PDF_fac}, which can significantly outperform the CD method adopted in \cite{Teng_2023_JSAC} when the user number is large. In each outer iteration of the HVMP algorithm, we perform multiple inner iterations of VMP to update these factors until convergence.
In the following, we shall introduce the updating rules for these factors sequentially. For convenience, the indices of both outer and inner iterations are omitted.
First, maximizing $\mathcal{L}(\chi_{\pmb{\vartheta},\pmb{W},\pmb{\alpha}|\pmb{Y}})$ w.r.t. the factor $\chi_{\pmb{\vartheta}|\pmb{Y}}(\pmb{\vartheta}^{\mathsf{UB}}_{k,g,t}|\mathbf{Y}_t)$ gives
\begin{align}\label{equ:opt_BCD_variable_vartheta_UB}
    &\ln \chi_{\pmb{\vartheta}|\pmb{Y}}(\pmb{\vartheta}^{\mathsf{UB}}_{k,g,t}|\mathbf{Y}_t) =  \ln \varepsilon_{\dot{f}_{k,t} \rightarrow \nu^{\mathsf{UB}}_{k,g,t}} + \ln \varepsilon_{f^{\mathsf{UB}}_{k,g,t} \rightarrow \theta^{\mathsf{UB}}_{k,g,t}} \notag \\
    &+ \ln \varepsilon_{f^{\mathsf{UB}}_{k,g,t} \rightarrow \tau^{\mathsf{UB}}_{k,g,t}} + \Re\{ (\pmb{\eta}^{\mathsf{UB}}_{k,g,t})^\mathsf{H} \mathbf{a}^{\mathsf{UB}}_{k,g,t} \},
\end{align}
where
\begin{small}
\begin{align}\label{equ:eta_def}
  &\pmb{\eta}^{\mathsf{UB}}_{k,g,t} = \frac{2}{\sigma^2_z}\bigg( (\hat{w}^{\mathsf{UB}}_{k,g,t})^{*}\mathbf{y}_{g,t} - \sum_{k' \ne k} \mathcal{M}_2(w^{\mathsf{UB}}_{k,g,t},w^{\mathsf{UB}}_{k',g,t})\mathbf{a}^{\mathsf{UB}}_{k',g,t}  \notag \\
  &-  \sum_{k''=1}^{K} \sum_{r=1}^{R} \mathcal{M}_2(w^{\mathsf{UB}}_{k,g,t},w^{\mathsf{UI}}_{k'',g,r,t}) \mathbf{a}^{\mathsf{UI}}_{k'',g,r,t} \bigg).
\end{align}
\end{small}In \eqref{equ:eta_def}, $\mathcal{M}_2(X,Y) = \mathbb{E}_{XY}[X^\mathsf{H} Y]$ denotes the expectation of $X^\mathsf{H}Y$ over the updated PDF in the last inner iteration. Similarly, maximizing $\mathcal{L}(\chi_{\pmb{\vartheta},\pmb{W},\pmb{\alpha}|\pmb{Y}})$ w.r.t. the factor $\chi_{\pmb{\vartheta}|\pmb{Y}}(\pmb{\vartheta}^{\mathsf{UI}}_{k,r,t}|\mathbf{Y}_t)$ also gives
\begin{align}\label{equ:opt_BCD_variable_vartheta_UI}
    &\ln \chi_{\pmb{\vartheta}|\pmb{Y}}(\pmb{\vartheta}^{\mathsf{UI}}_{k,r,t}|\mathbf{Y}_t) =  \ln \varepsilon_{\dot{f}_{k,t} \rightarrow \nu^{\mathsf{UI}}_{k,g,t}} + \ln \varepsilon_{f^{\mathsf{UI}}_{k,g,t} \rightarrow \theta^{\mathsf{UI}}_{k,g,t}} \notag \\
    &+ \ln \varepsilon_{f^{\mathsf{UI}}_{k,g,t} \rightarrow \tau^{\mathsf{UI}}_{k,g,t}} + \sum_{g=1}^{G}\Re\{ (\pmb{\eta}^{\mathsf{UI}}_{k,g,r,t})^\mathsf{H} \mathbf{a}^{\mathsf{UI}}_{k,g,r,t} \} \notag \\
    &- \Re\{(\tilde{\pmb{\eta}}^{\mathsf{UI}}_{k,r,t})^\mathsf{H} \mathbf{a}_{\mathsf{I}}(\theta^{\mathsf{UI}}_{k,r,t})\},
\end{align}
where
\begin{small}
\begin{align}
  &\pmb{\eta}^{\mathsf{UI}}_{k,g,r,t} = \frac{2}{\sigma^2_z} \bigg(  - \sum_{k'=1}^{K} \mathcal{M}_2(w^{\mathsf{UI}}_{k,g,r,t},w^{\mathsf{UB}}_{k',g,t})\mathbf{a}^{\mathsf{UB}}_{k',g,t}  \notag \\
  &-  \sum_{k'' \ne k} \sum_{r=1}^{R} \mathcal{M}_2(w^{\mathsf{UI}}_{k,g,r,t},w^{\mathsf{UI}}_{k'',g,r,t}) \mathbf{a}^{\mathsf{UI}}_{k'',g,r,t} + (\hat{w}^{\mathsf{UI}}_{k,g,r,t})^{*}\mathbf{y}_{g,t} \bigg), \notag 
\end{align}
\end{small}and each $m_{\mathsf{I}}$-th element in $\tilde{\pmb{\eta}}^{\mathsf{UI}}_{k,r,t}$ is obtained by \eqref{equ:eta_beam} at the top of the next page.
\begin{figure*}
\begin{small}
\begin{align}\label{equ:eta_beam}
  [\tilde{\pmb{\eta}}^{\mathsf{UI}}_{k,r,t}]_{m_{\mathsf{I}}} =  \left\{ \begin{array}{ll}
                                                      \frac{M_{\mathsf{B}}D_{\mathsf{I}}QM_{\mathsf{I}}}{\sigma^2_z} \sum_{g=1}^{G}\mathcal{M}_2(w^{\mathsf{UI}}_{k,g,r,t},w^{\mathsf{UI}}_{k,g,r,t}), & m_{\mathsf{I}} = 1, \\
                                                      \frac{2M_{\mathsf{B}}D_{\mathsf{I}} Q_2 }{\sigma^2_z} \sum_{g=1}^{G} \sum_{i = m_{\mathsf{I}}}^{M_{\mathsf{I}}}  \mathcal{M}_2(w^{\mathsf{UI}}_{k,g,r,t},w^{\mathsf{UI}}_{k,g,r,t})[\bar{\pmb{\Psi}}_{r,t}^{*}\bar{\pmb{\Psi}}_{r,t}^{T}]_{i,i-m_{\mathsf{I}}+1}, & m_{\mathsf{I}} \ne 1,
                                                    \end{array}\right.
\end{align}
\end{small}
\hrule
\end{figure*}
Note that the last term exists in \eqref{equ:opt_BCD_variable_vartheta_UI} due to the beamspace signal structure by the cascade channel of the user-RIS-BS links. Moreover, the variable vector $\pmb{\vartheta}^{\mathsf{UB}}_{k,g,t}$ is only statistically dependent on the received signal at the BS $g$, while $\pmb{\vartheta}^{\mathsf{UI}}_{k,r,t}$ is statistically dependent on the received signal at all BSs, which contributes to the hierarchical signal processing operation. 
The variable vector $\pmb{\vartheta}^{\mathsf{UB}}_{k,g,t}$ in \eqref{equ:opt_BCD_variable_vartheta_UB} is then refined via the Newton method \cite{Zhang_2020_SPL} as 
\begin{align}
  \hat{\pmb{\vartheta}}^{\mathsf{UB}}_{k,g,t} = \hat{\pmb{\vartheta}}^{\mathsf{UB}, \mathsf{old}}_{k,g,t} - ( \pmb{\triangledown}^2 \mathcal{G}^{\mathsf{B}}(\hat{\pmb{\vartheta}}^{\mathsf{UB}, \mathsf{old}}_{k,g,t}) )^{-1} \pmb{\triangledown} \mathcal{G}^{\mathsf{B}}(\hat{\pmb{\vartheta}}^{\mathsf{UB}, \mathsf{old}}_{k,g,t}), 
\end{align}
where $\mathcal{G}^{\mathsf{B}}(\pmb{\vartheta}^{\mathsf{UB}}_{k,g,t})$ is defined as \eqref{equ:opt_BCD_variable_vartheta_UB}, and $\pmb{\triangledown} \mathcal{G}^{\mathsf{B}}(\pmb{\vartheta}^{\mathsf{UB}, \mathsf{old}}_{k,g,t})$ ($\pmb{\triangledown}^2 \mathcal{G}^{\mathsf{B}}(\pmb{\vartheta}^{\mathsf{UB}, \mathsf{old}}_{k,g,t}))$ denotes the gradient (Hessen matrix) at $\pmb{\vartheta}^{\mathsf{UB}, \mathsf{old}}_{k,g,t}$. Similarly, the Newton method is also utilized to refine the estimates of $\pmb{\vartheta}^{\mathsf{UI}}_{k,r,t}$ with \eqref{equ:opt_BCD_variable_vartheta_UI}.
Then, the concentration parameter vectors $\hat{\pmb{\kappa}}^{\mathsf{UB}}_{k,g,t} = [\hat{\kappa}^{\mathsf{UB}}_{k,g,t}(\theta),\hat{\kappa}^{\mathsf{UB}}_{k,g,t}(\tau),\hat{\kappa}^{\mathsf{UB}}_{k,g,t}(\nu)]^\mathsf{T}$ and $\hat{\pmb{\kappa}}^{\mathsf{UI}}_{k,r,t} = [\hat{\kappa}^{\mathsf{UI}}_{k,r,t}(\theta),\hat{\kappa}^{\mathsf{UI}}_{k,r,t}(\tau),\hat{\kappa}^{\mathsf{UI}}_{k,r,t}(\nu)]^\mathsf{T}$ can also be updated by following \cite{Zhang_2020_SPL}.

Next, maximizing $\mathcal{L}(\chi_{\pmb{\vartheta},\pmb{W},\pmb{\alpha}|\pmb{Y}})$ w.r.t. the factor $\chi_{\pmb{W}|\pmb{Y}}(\mathbf{w}_{g,t}|\mathbf{Y}_t)$ gives
\begin{align}\label{equ:opt_BCD_w}
  \chi_{\pmb{W}|\pmb{Y}}(\mathbf{w}_{g,t}|\mathbf{Y}_t) =&~ \mathcal{CN}([\mathbf{w}_{g,t}]_{\hat{\mathcal{A}}_{g,t}};[\hat{\mathbf{w}}_{g,t}]_{\hat{\mathcal{A}}_{g,t}},[\hat{\mathbf{C}}_{g,t}]_{\hat{\mathcal{A}}_{g,t},\hat{\mathcal{A}}_{g,t}}) \notag \\
  &\times \prod_{j \notin \hat{\mathcal{A}}} \delta([\mathbf{w}_{g,t}]_{j}),
\end{align}
where $\hat{\mathcal{A}}_{g,t}$ is the support set of the vector $\hat{\mathbf{w}}_{g,t}$ and $[\hat{\mathbf{w}}_{g,t}]_{\hat{\mathcal{A}}_{g,t}}$ is the vector containing the elements in the subset $\hat{\mathcal{A}}_{g,t}$. In \eqref{equ:opt_BCD_w}, the mean and covariance of $[\mathbf{w}_{g,t}]_{\hat{\mathcal{A}}_{g,t}}$ is given by
\begin{align}
  [\hat{\mathbf{w}}_{g,t}]_{\hat{\mathcal{A}}_{g,t}} &= \sigma^{-2}_{z} [\hat{\mathbf{C}}_{g,t}]_{\hat{\mathcal{A}}_{g,t},\hat{\mathcal{A}}_{g,t}} [\hat{\mathbf{A}}_{g,t}]^\mathsf{H}_{:,\hat{\mathcal{A}}_{g,t}}\mathbf{y}_{g,t}, \notag \\
  [\hat{\mathbf{C}}_{g,t}]_{\hat{\mathcal{A}}_{g,t},\hat{\mathcal{A}}_{g,t}} &= \sigma^{2}_{z}([\mathbf{J}_{g,t}]_{\hat{\mathcal{A}}_{g,t},\hat{\mathcal{A}}_{g,t}} + \sigma^{2}_{z}\text{diag}([\breve{\pmb{\beta}}_{g,t}]_{\hat{\mathcal{A}}_{g,t}}))^{-1}, \notag 
\end{align}
where $[\breve{\pmb{\beta}}_{g,t}]_i = [\pmb{\beta}_{g,t}]^{-1}_i, ~\forall i$ and $\mathbf{J}_{g,t} = \mathbb{E}[\mathbf{A}^\mathsf{H}_{g,t}\mathbf{A}_{g,t}]$.

Since $\chi_{\pmb{\alpha}|\pmb{Y}}([\pmb{\alpha}_{g,t}]_{i}|\mathbf{Y}_t) = \delta([\pmb{\alpha}_{g,t}]_{i} - [\hat{\pmb{\alpha}}_{g,t}]_{i})$, we propose to determine $[\hat{\pmb{\alpha}}_{g,t}]$ via hypothesis test given by
\begin{align}\label{equ:LLR_test}
  \log \left( \frac{p([\hat{\mathbf{w}}_{g,t}]_{i}|[\hat{\pmb{\alpha}}_{g,t}]_{i} = 1)}{p([\hat{\mathbf{w}}_{g,t}]_{i}|[\hat{\pmb{\alpha}}_{g,t}]_{i} = 0)} \right) \mathop{\gtrless}\limits_{H_0}^{H_1} \gamma_i,
\end{align}
where $\gamma_i$ is the pre-determined threshold obtained by empirical results. Compared with the greedy iterative search method in \cite{Badiu_2017_TSP}, the hypothesis test-based method can determine all of these indicator variables in parallel without an iterative searching process to find the local optimal solution and is thus computationally efficient in large-scale systems. Based on \eqref{equ:LLR_test}, the support set $\hat{\mathcal{A}}_{g,t}$ can also be updated. By iteratively executing the above calculation procedure, the estimation of $\{\pmb{\vartheta}_t,\mathbf{W}_t,\pmb{\alpha}_t\}$ can be refined.
After the VMP algorithm converges, we adopt the variable $\epsilon_{\theta^{\mathsf{UB}}_{k,g,t}}$ as an example to show the derivation of the extrinsic message by
\begin{align}
  \varepsilon_{\theta^{\mathsf{UB}}_{k,g,t} \leftarrow py_t} = \mathcal{VM}(\theta^{\mathsf{UB}}_{k,g,t}; \overleftarrow{\mu}^{\mathsf{UB}}_{k,g,t}(\theta), \overleftarrow{\kappa}^{\mathsf{UB}}_{k,g,t}(\theta)),
\end{align}
where
\begin{align}
  \overleftarrow{\kappa}^{\mathsf{UB}}_{k,g,t}(\theta) e^{\jmath \overleftarrow{\mu}^{\mathsf{UB}}_{k,g,t}(\theta)} =&~ \hat{\kappa}^{\mathsf{UB}}_{k,g,t}(\theta) e^{- \jmath \zeta_{\mathsf{S}}\hat{\theta}^{\mathsf{UB}}_{k,g,t}} \notag \\
  &- \overrightarrow{\kappa}^{\mathsf{UB}}_{k,g,t}(\theta) e^{ \jmath \overrightarrow{\mu}^{\mathsf{UB}}_{k,g,t}(\theta)},
\end{align}
since the family of VM distributions is closed under the multiplication and division operations. Similarly, extrinsic messages on the other position-related parameters can also be derived.

\subsubsection{Forward messages}
After the state estimation in the current time slot $t$ via the above bidirectional message passing is converged, we can update the forward messages to the time slot $t+1$ for online user tracking and signal detection. These forward messages can be expressed by
\begin{align}
    \varepsilon_{\dot{u}_{k,t+1} \rightarrow \dot{\mathbf{p}}_{k,t+1}} &= \varepsilon_{\dot{\mathbf{p}}_{k,t} \rightarrow \dot{u}_{k,t+1}} = \varepsilon_{\dot{\mathbf{p}} \rightarrow u_{k,t+1}} \notag \\
    &\varpropto \varepsilon_{\dot{u}_{k,t} \rightarrow \dot{\mathbf{p}}_{k,t}}(\infty) \varepsilon_{\dot{\mathbf{p}}_{k,t} \leftarrow \dot{f}_{k,t}}(\infty) \notag \\
    &= \mathcal{N}(\dot{\mathbf{p}}_{k,t+1};\dot{\overrightarrow{\mathbf{p}}}_{k,t+1},\dot{\overrightarrow{\mathbf{Q}}}_{k,t+1}), \label{equ:mp_forward_vel} \\
    \varepsilon_{\mathbf{p}_{k,t} \rightarrow u_{k,t+1}} &\varpropto \varepsilon_{u_{k,t} \rightarrow \mathbf{p}_{k,t}} \varepsilon_{\mathbf{p}_{k,t} \leftarrow f_{k,t}}(\infty) \notag \\
    &= \mathcal{N}(\mathbf{p}_{k,t+1};\hat{\mathbf{p}}_{k,t+1},\hat{\mathbf{Q}}_{k,t+1}), \label{equ:mp_fusion_pos}   \\
    \varepsilon_{u_{k,t+1} \rightarrow \mathbf{p}_{k,t+1}} &\varpropto \varepsilon_{\dot{\mathbf{p}} \rightarrow u_{k,t+1}} \varepsilon_{\mathbf{p}_{k,t} \rightarrow u_{k,t+1}} \notag \\
    &= \mathcal{N}(\mathbf{p}_{k,t+1};\overrightarrow{\mathbf{p}}_{k,t+1},\overrightarrow{\mathbf{Q}}_{k,t+1}). \label{equ:mp_forward_pos}
\end{align}
Since all the above messages in the derivation of \eqref{equ:mp_forward_vel} -- \eqref{equ:mp_forward_pos} are in the family of Gaussian distributions, explicit expressions of these messages for velocities and positions are easy to be obtained and thus omitted here. 

\begin{algorithm}[t]
    \small
	\caption{The HVMP Algorithm.}\label{alg:HVMP}
    {\bf Input}: $p(\pmb{\psi}_0)$ and the received signals $\{\mathbf{Y}_t\}$;\\
    {\bf Output}: States $\{\hat{\pmb{\psi}}_{t}\}$ and signals $\{\hat{\mathbf{s}}_t\}$;

    \begin{algorithmic}[1]  
    \FOR {$t = 1$ to $V$} 
        \FOR{$j = 1$ to $J_{\rm out}$}
        \STATE Update $\varepsilon_{\dot{\mathbf{p}}_{k,t} \rightarrow \dot{f}_{k,t}}(j)$ and $\varepsilon_{\mathbf{p}_{k,t} \rightarrow f_{k,t}}(j)$ in \eqref{equ:m_dp_to_df} and \eqref{equ:m_p_to_f}, respectively;
        \STATE Update $\varepsilon_{f_{k,t} \rightarrow \mathbf{p}^{\mathsf{UB}}_{k,g,t}}(j)$/$\varepsilon_{f_{k,t} \rightarrow \mathbf{p}^{\mathsf{UI}}_{k,r,t}}(j)$ in \eqref{equ:m_pUB_bto_f};
        \STATE Update $\varepsilon_{\mathbf{p}^{\mathsf{UB}}_{k,g,t} \rightarrow f^{\mathsf{UB}}_{k,g,t}}(j) = \varepsilon_{f_{k,t} \rightarrow \mathbf{p}^{\mathsf{UB}}_{k,g,t}}(j)$ and $\varepsilon_{\mathbf{p}^{\mathsf{UI}}_{k,r,t} \rightarrow f^{\mathsf{UI}}_{k,r,t}}(j) = \varepsilon_{f_{k,t} \rightarrow \mathbf{p}^{\mathsf{UI}}_{k,r,t}}(j)$;
        \STATE Update $\varepsilon_{\dot{f}_{k,t} \rightarrow \nu^{\mathsf{UB}}_{k,g,t}}(j)$/$\varepsilon_{\dot{f}_{k,t} \rightarrow \nu^{\mathsf{UI}}_{k,r,t}}(j)$ in \eqref{equ:m_df_to_nu};
        \STATE Update $\varepsilon_{f^{\mathsf{UB}}_{k,g,t} \rightarrow \theta^{\mathsf{UB}}_{k,g,t}}(j)$/$\varepsilon_{f^{\mathsf{UI}}_{k,r,t} \rightarrow \theta^{\mathsf{UI}}_{k,r,t}}(j)$ similar to \eqref{equ:m_df_to_nu};
        \STATE Update $\varepsilon_{f^{\mathsf{UB}}_{k,g,t} \rightarrow \tau^{\mathsf{UB}}_{k,g,t}}(j)$/$\varepsilon_{f^{\mathsf{UI}}_{k,r,t} \rightarrow \tau^{\mathsf{UI}}_{k,r,t}}(j)$ similar to \eqref{equ:m_df_to_nu};
        \STATE Update $\varepsilon_{l_{k,t} \rightarrow w^{\mathsf{UB}}_{k,g,t}}(j)$/$\varepsilon_{l_{k,t} \rightarrow w^{\mathsf{UI}}_{k,r,t}}(j)$ in \eqref{equ:l_r_w};
        \STATE Perform VMP in Sec. \ref{sec:VMP}) (with maximal $J_{\rm inn}$ inner iterations) to update $\varepsilon_{\theta^{\mathsf{UB}}_{k,g,t} \leftarrow py_t}$/$\varepsilon_{\theta^{\mathsf{UI}}_{k,r,t} \leftarrow py_t}(j)$, $\varepsilon_{\tau^{\mathsf{UB}}_{k,g,t} \leftarrow py_t}(j)$/$\varepsilon_{\tau^{\mathsf{UI}}_{k,r,t} \leftarrow py_t}(j)$, $\varepsilon_{\nu^{\mathsf{UB}}_{k,g,t} \leftarrow py_t}(j)$/$\varepsilon_{\nu^{\mathsf{UI}}_{k,r,t} \leftarrow py_t}(j)$, and $\varepsilon_{w^{\mathsf{UB}}_{k,g,t} \leftarrow py_t}(j)$/$\varepsilon_{w^{\mathsf{UI}}_{k,r,t} \leftarrow py_t}(j)$;
        \STATE Update $\varepsilon_{\dot{f}_{k,t} \leftarrow \nu^{\mathsf{UB}}_{k,g,t}}(j)$/$\varepsilon_{\dot{f}_{k,t} \leftarrow \nu^{\mathsf{UI}}_{k,r,t}}(j)$ in \eqref{equ:m_nu_to_df};
        \STATE Update $\varepsilon_{f^{\mathsf{UB}}_{k,g,t} \leftarrow \theta^{\mathsf{UB}}_{k,g,t}}(j)$/$\varepsilon_{f^{\mathsf{UI}}_{k,r,t} \leftarrow \theta^{\mathsf{UI}}_{k,r,t}}(j)$ similar to \eqref{equ:m_nu_to_df};
        \STATE Update $\varepsilon_{f^{\mathsf{UB}}_{k,g,t} \leftarrow \tau^{\mathsf{UB}}_{k,g,t}}(j)$/$\varepsilon_{f^{\mathsf{UI}}_{k,r,t} \leftarrow \tau^{\mathsf{UI}}_{k,r,t}}(j)$ similar to \eqref{equ:m_nu_to_df};
        \STATE Update $\varepsilon_{l_{k,t} \leftarrow w^{\mathsf{UB}}_{k,g,t}}(j)$/$\varepsilon_{l_{k,t} \leftarrow w^{\mathsf{UI}}_{k,r,t}}(j)$ in \eqref{equ:l_l_w};
        \STATE Update $\varepsilon_{f_{k,t} \leftarrow \mathbf{p}^{\mathsf{UB}}_{k,g,t}}(j)$/$\varepsilon_{f_{k,t} \leftarrow \mathbf{p}^{\mathsf{UI}}_{k,r,t}}(j)$ in \eqref{equ:m_pUB_to_f};
        \STATE Update $\varepsilon_{\dot{\mathbf{p}}_{k,t} \leftarrow \dot{f}_{k,t}}(j)$, $\varepsilon_{\mathbf{p}_{k,t} \leftarrow f_{k,t}}(j)$ , and $\varepsilon_{\tilde{s}_{k,t} \leftarrow l_{k,t}}(j)$ in \eqref{equ:m_df_to_dp}, \eqref{equ:mp_fusion_p}, and \eqref{equ:m_s_bto_l}, respectively;
        \ENDFOR
        \STATE Update $\varepsilon_{\dot{u}_{k,t+1} \rightarrow \dot{\mathbf{p}}_{k,t+1}}$ in \eqref{equ:mp_forward_vel};
        \STATE Update $\varepsilon_{\mathbf{p}_{k,t} \rightarrow u_{k,t+1}}$ in \eqref{equ:mp_fusion_pos};
        \STATE Update $\varepsilon_{u_{k,t+1} \rightarrow \mathbf{p}_{k,t+1}}$ in \eqref{equ:mp_forward_pos};
    \ENDFOR
    \end{algorithmic} 
\end{algorithm}

In summary, we iteratively perform bidirectional message passing to update the states and position-related parameters based on the prior information from the time slot $t-1$ until HVMP converges, and then utilize the converged estimation results to refine the prior information of user states in the next time slot $t+1$ via forward message passing, which can realize the online tracking and signal detection for all users with diverse and high mobility. The whole algorithm is outlined in Algorithm \ref{alg:HVMP}.

\begin{remark}
It is highlighted that the proposed HVMP algorithm leverages MP techniques to effectively combine measurements from LOS and reflecting links, where the blockage conditions of different links, i.e., $\{\pmb{\alpha}_{g,t}\}$, are automatically determined during each iteration. Therefore, the HVMP algorithm has superior flexibility and accuracy by iteratively estimating the blockage condition of each link during the parameter estimation procedure.
\end{remark}

\begin{remark}
By considering BS cooperation, the fronthual links for aggregating received signals at each BS may have finite capacity due to the hardware limit \cite{Zhu_2024_TWC}. Thus, the quantize-and-forward (QF) scheme \cite{Zhu_2024_TWC} can be adopted and the proposed algorithm can be applied to solve the considered problem by combining the dequantization technique in \cite{Zhu_2024_TWC}. However, the utilized quantization scheme also has a large influence on the joint estimation performance, which is out of the scope for this work and can be left as future works.
\end{remark}


\section{Performance Analysis and Optimization}

\subsection{Performance Analysis via Bayesian Cramér-Rao Bound}
In this work, we adopt the Bayesian Cramér-Rao Bound to characterize the tracking performance of users, which is widely used in the estimation problem with random parameters.

We define the estimated parameter vector $\tilde{\pmb{\psi}}_t = [\pmb{\psi}_{t}^\mathsf{T}, \tilde{\mathbf{s}}_t^\mathsf{T}]^\mathsf{T} \in \mathbb{C}^{5K \times 1}$. 
Then, the Bayesian information matrix (BIM) for the discrete-time tracking nonlinear filtering method \cite{Tichavsky_1998_TSP} can be given by the following recursive derivation of
\begin{align}\label{equ:BIM}
    \mathbf{B}_t = \mathbf{B}^{\rm mea}_{t} + \pmb{\Xi}^{22}_{t} - \pmb{\Xi}^{21}_{t}( \mathbf{B}_{t-1} + \pmb{\Xi}^{11}_{t})^{-1} \pmb{\Xi}^{12}_{t},
\end{align}
where
\begin{small}
\begin{align}
  &\mathbf{B}^{\rm mea}_{t}  = \mathbb{E}_{\mathbf{y}, \tilde{\pmb{\psi}}_t}\left[ -\frac{\partial^2 \log p(\mathbf{y}_t|\tilde{\pmb{\psi}}_t)}{\partial \tilde{\pmb{\psi}}_t \partial \tilde{\pmb{\psi}}_t^\mathsf{H}} \right], \label{equ:FIM_data} \\
  &\pmb{\Xi}^{ij}_{t} = \mathbb{E}_{\tilde{\pmb{\psi}}_t,\tilde{\pmb{\psi}}_{t-1}} \left[ - \frac{\partial^2 \log p(\tilde{\pmb{\psi}}_t|\tilde{\pmb{\psi}}_{t-1})}{\partial \tilde{\pmb{\psi}}_{t+i-2} \partial \tilde{\pmb{\psi}}_{t+j-2}^\mathsf{H}} \right],  ~\forall i,j \in \{1,2\}.
\end{align}
\end{small}The detailed derivation of the measurement BIM $\mathbf{B}^{\rm mea}_{t}$ are omitted here due to page limits. 

For the initialization of the BCRB calculation, we have $\mathbf{B}_{0} = \mathbb{E}_{\tilde{\pmb{\psi}}_{0}} \left[ -\frac{\partial^2 \log p(\tilde{\pmb{\psi}}_{0})}{\partial \tilde{\pmb{\psi}}_{0} \partial \tilde{\pmb{\psi}}_{0}^\mathsf{H}}  \right]$. Then, for each tracking time slot $t$, we can obtain the estimation result $\bar{\pmb{\psi}}_t$ satisfies
\begin{align}
  \mathbb{E}\left[ (\widehat{\pmb{\psi}}_t - \tilde{\pmb{\psi}}_t) (\widehat{\pmb{\psi}}_t - \tilde{\pmb{\psi}}_t)^\mathsf{H} \right] \succeq \mathbf{B}^{-1}_t, ~\forall t.
\end{align}
Therefore, the BCRB serves as the lower bound for the parameter estimation performance, which is potential to help optimize the RIS phase profile during the tracking procedure.

\subsection{RIS Phase Profile Optimization}

During the tracking procedure, the RIS phase profile can be adaptively optimized to realize high-accuracy tracking and signal detection performance. As introduced in the last subsection, the BCRB can be leveraged as the performance metric for the state estimation performance and the signal detection performance. As such, we can formulate an optimization problem with the objective function of the BCRB over the variable of the RIS phase profile $\{\varsigma^{(q)}_{m_{\mathsf{I}},r} = \angle \phi^{(q)}_{m_{\mathsf{I}},r}\}$. In contrast to directly minimizing the BCRB, we consider to minimize the general weighted BCRB and the optimization problem can be expressed by
\begin{align}\label{equ:Opt_phase}
  (\text{P1}) \quad \min_{\{\pmb{\varsigma}_r\}} &~~ \text{Trace}\{ \text{diag}(\mathbf{v}) \mathbf{B}^{-1}_t \},
\end{align} 
where $\mathbf{v} = [v_{1},\dots,v_{5K}]^\mathsf{T}$ is the vector to adjust the weight on each estimated parameter.\footnote{For example, we may only care about the estimation accuracy of the user positions and transmit signals, meaning that we can set the weights related to velocities to be zeros.} 
To solve Problem (P1), we leverage the gradient-based algorithm that can obtain a local optimal solution due to the non-convex objective function. 
In detail, in each iteration $e$, the updating rule for the RIS phase profile is given by
\begin{align}
  \pmb{\varsigma}_r^{(e)} = \pmb{\varsigma}_r^{(e-1)} - \aleph^{(e)} \pmb{\triangledown} \mathcal{F}(\{\pmb{\varsigma}_{r}\}), \forall e, r,
\end{align}
where the step size $\aleph^{(e)}$ is determined by the Armijo backtracking line search. The gradient-based method will iteratively update the RIS phase profile until the stopping criterion $\sum_{r=1}^{R}||\aleph^{(e)} \pmb{\triangledown}_r \mathcal{F}(\{\pmb{\varsigma}_{r}\})|| < \epsilon$ is satisfied.

\section{Simulation Results}\label{sec:simu}

This section provides simulation results to verify the effectiveness of the proposed algorithm for the RIS-assisted ISAC system. In the system setup, we consider that there are $G=2$ BSs deployed at $\mathbf{p}^{\mathsf{B}}_1 = [0,0]^\mathsf{T}$ m and $\mathbf{p}^{\mathsf{B}}_2 = [90,0]^\mathsf{T}$ m, $R=2$ RISs deployed at $\mathbf{p}^{\mathsf{I}}_{1} = [20,40]^\mathsf{T}$ m and $\mathbf{p}^{\mathsf{I}}_{2} = [60,40]^\mathsf{T}$ m, and $K = 3$ users, which move in the square region with the center $[45,-5]^{\mathsf{T}}$ m and the side length $50$ m. We also consider that average velocities of these three users are $\bar{\upsilon}_{1} = 40$ m/s, $\bar{\upsilon}_2 = 30$ m/s, $\bar{\upsilon}_3 = 15$ m/s, where $\bar{\upsilon}_{k} = T^{-1}\sum_{t=1}^{T}||\dot{\mathbf{p}}^{\mathsf{U}}_{k,t}||_2, ~\forall k$. Each BS is equipped with $M_{\mathsf{B}} = 6$ antennas and each RIS is equipped with $M_{\mathsf{I}} = 48$ elements. We consider that the bandwidth is $N\Delta f = 10$ MHz with $N = 12$ subcarriers and $\mathcal{N}_{0} = \{1,2,\dots,12\}$ with $\Delta N = 1$. The tracking interval is set as $\Delta T = 0.02$ s and the ISAC sub-block is set with $Q_1 = 12$, $I = 10$, and $\Delta Q = 200$. The path loss factor is modeled by $\beta = \frac{\lambda}{4\pi d}$ \cite{Teng_2023_JSAC,Zhu_2026_TWC} for each LOS link. We consider that each user transmits Gaussian signals if not specified. All the simulation results are averaged over $100$ channel realizations. The tracking performance is evaluated by the metric of the root MSE (RMSE): $\text{RMSE} = T^{-1}\sum_{t=1}^{T}||\hat{\pmb{\psi}}_{t} - \pmb{\psi}_{t}||_2$, and the signal detection performance is evaluated by the MSE: $\text{MSE} = T^{-1}\sum_{t=1}^{T}||\hat{\mathbf{s}}_t - \tilde{\mathbf{s}}_t||_2^2$ \cite{Guo_2005_TIT}. For comparison, we adopt three representative methods as benchmarks: 

1) \textbf{Benchmark 1:} MUSIC with spatial smoothing \cite{Shan_1985_TASSP} is utilized to estimate the user states from the received signal $\mathbf{Y}_{t}$ and then KF is applied to refine the state estimation; 

2) \textbf{Benchmark 2:} the variational inference algorithm only tracks the position of each user without Doppler frequency (velocity) estimation, which is similar to that in \cite{Teng_2023_JSAC}; 

3) \textbf{Benchmark 3:} we consider to perform multi-user tracking under the pilot-based transmission protocol without performing signal detection in the pilot phase, which can serve as one lower bound on multi-user tracking.

\begin{figure}[t]
  \centering
  \includegraphics[width=.48\textwidth]{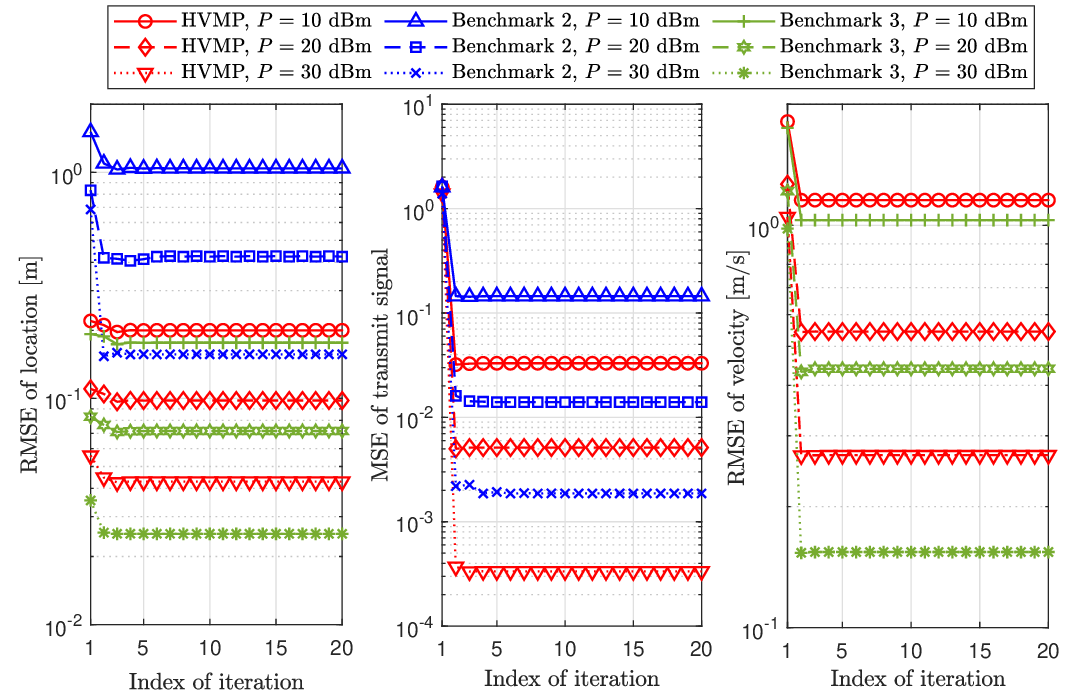}
  \vspace{-0.3cm}
  \caption{The convergence performance of the proposed algorithm: (left) position estimation, (middle) signal detection, (right) velocity estimation.}\label{Fig:RMSE_iter}
  \vspace{-0.3cm}
\end{figure}

First, we evaluate the convergence performance in Fig. \ref{Fig:RMSE_iter} under different transmit powers. We can observe that the proposed HVMP algorithm can quickly converge to the stationary solutions within several outer iterations. Compared with Benchmark 2, the HVMP algorithm can consistently provide significantly smaller estimation errors for both states and transmit signals. In particular, it is observed that the initial estimation error of HVMP, known as the location prediction error, is also small and approaching the converged estimation error, indicating the effectiveness of exploiting the full state of each user for tracking performance enhancement. It is also validated that Benchmark 3 performs as a lower bound for the state estimation of the proposed scheme, which utilize pilot signals to enhance the parameter estimation performance.
In the following, we only evaluate the performance of the location and transmit signal estimation, which act as the key performance metrics for our considered problem.

\begin{figure}[t]
  \centering
  \includegraphics[width=.48\textwidth]{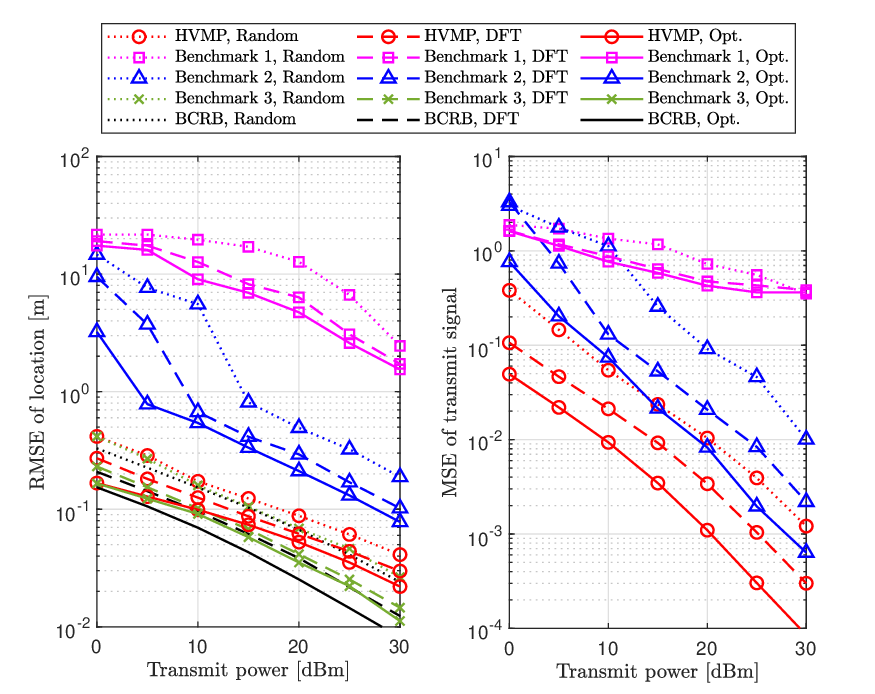}
  \vspace{-0.5cm}
  \caption{The performance comparison under different transmit powers: (left) position estimation, (right) signal detection.}\label{Fig:RMSE_SNR}
  \vspace{-0.5cm}
\end{figure}

We then evaluate the impact of the RIS phase profile design on the performance of the proposed algorithm and benchmarks in Fig. \ref{Fig:RMSE_SNR}. Three kinds of RIS phase profile design are considered: random phase profile, DFT phase profile, and optimized phase profile. For the DFT phase profile, we consider that the $\lfloor \frac{Q_1}{K} \rfloor$ DFT codewords with the closest distances to $\tilde{\mathbf{a}}_{\mathsf{I}}(\tilde{\theta}^{\rm UI}_{k,t})$ are selected for each user $k$ at each RIS $r$, where $\tilde{\theta}^{\rm UI}_{k,r,t}$ is the predicted AOA from user $k$ to the RIS $r$. We can find that all of the proposed algorithm and benchmarks can benefit from the optimized RIS phase profile to enhance the tracking and signal detection performance, validating the effectiveness of the proposed optimization method. Due to the elaborated updating rules and full exploitation of user states, the proposed HVMP algorithm can significantly outperform the two representative benchmarks and approach the performance limit of BCRB. Moreover, we can observe that Benchmark 2 can provide superior performance to Benchmark 1 even without Doppler frequency estimation, since Benchmark 1 via MUSIC suffers from poor performance on the state estimation in the case with inadequate time samples (snapshots).

\begin{figure}[t]
  \centering
  \includegraphics[width=.45\textwidth]{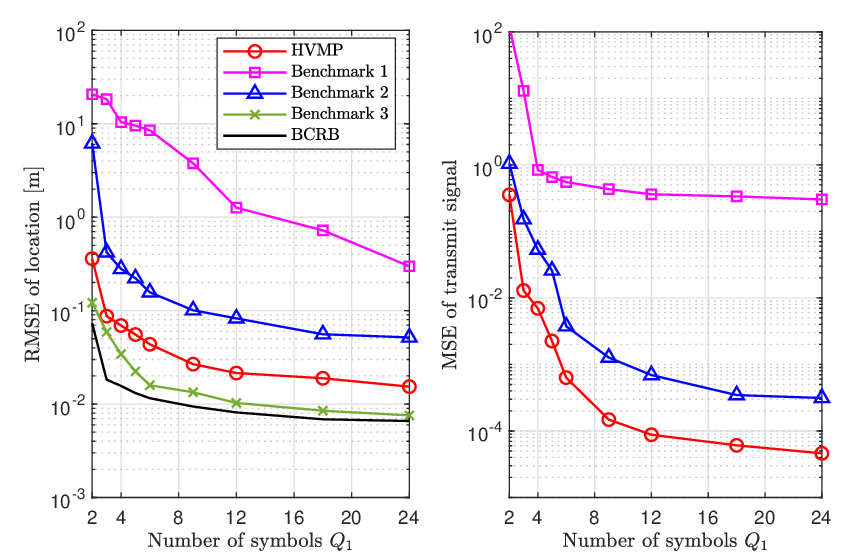}
  \vspace{-0.5cm}
  \caption{The performance comparison under different numbers of OFDM symbols $Q_1$ in each group of the ISAC sub-block with $P = 30$ dBm: (left) position estimation, (right) signal detection.}\label{Fig:RMSE_Q1}
  \vspace{-0.5cm}
\end{figure}

The performance of the proposed algorithm with different numbers of OFDM symbols $Q_1$ in each group of the ISAC sub-block is given in Fig. \ref{Fig:RMSE_Q1}. Note the condition $Q_1 \ge 2$ needs to be satisfied since we cannot extract the AOA information of user-RIS links with fixed RIS phase profile during each tracking time slot, i.e., $Q_1 = 1$. From Fig. \ref{Fig:RMSE_Q1}, the tracking and signal detection errors are first reduced rapidly by enlarging $Q_1$, since we can get much more accurate AOA information for each user-RIS link to enhance the location estimation performance. When $Q_1$ exceeds a certain value, the proposed algorithm turns to have saturated tracking performance, since the distance estimation errors dependent on the bandwidth become the limiting factor for the tracking performance at this moment. This result also reveals that we only require a moderate number of OFDM symbols to accomplish the high-precision tracking task, which can save the tracking overhead to promote the data transmission performance.

\begin{figure}[t]
  \centering
  \includegraphics[width=.45\textwidth]{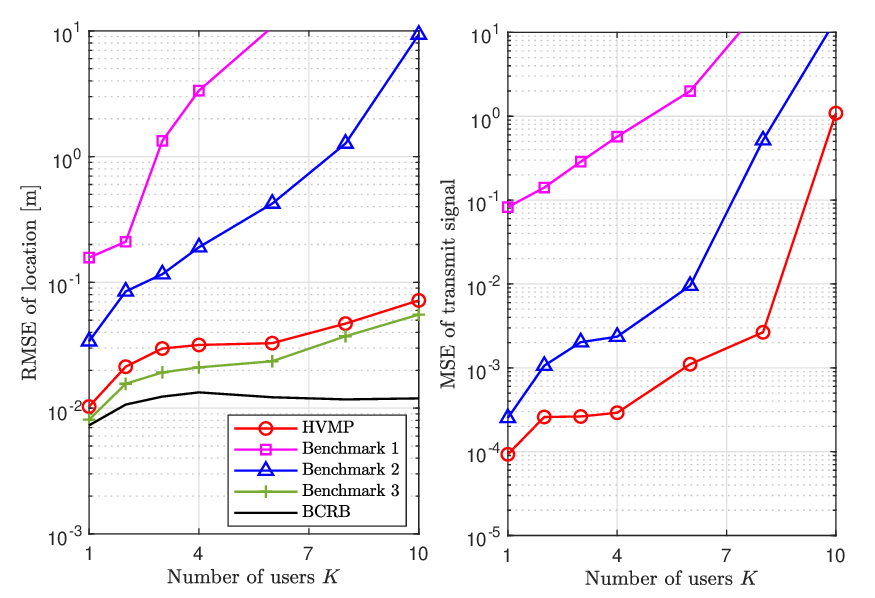}
  \vspace{-0.5cm}
  \caption{The performance comparison under different numbers of users $K$  with $P = 30$ dBm: (left) position estimation, (right) signal detection.}\label{Fig:RMSE_K}
  \vspace{-0.5cm}
\end{figure}

We also evaluate the performance of the proposed algorithm under different numbers of users in Fig. \ref{Fig:RMSE_K}, where the average velocities of different users are in the region $[10,40]$ m/s. With $K$ increasing, both of the two benchmarks are shown to provide significantly deteriorated tracking and signal detection performance and gradually fail to accurately track the locations of users. In contrast, the proposed HVMP algorithm suffers from much milder tracking performance degradation and can provide consistent centimeter-level tracking performance when $K \le 10$. These results verify the robust tracking performance of the proposed algorithm in the scenario with a wide range of user numbers, implying that the HVMP algorithm is suitable to the practical scenario with a varying number of users. Compared with the tracking performance, the proposed algorithm suffers from larger estimation error on the transmit signals due to more severe interference between users, indicating the necessity of increasing the tracking overhead to enhance both of the channel estimation and signal detection performance.

\begin{figure}[t]
  \centering
  \includegraphics[width=.48\textwidth]{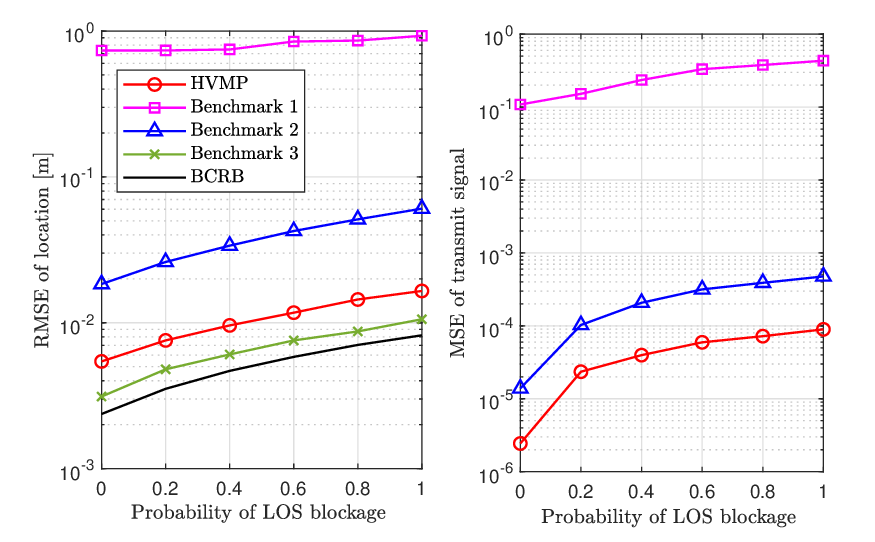}
  \vspace{-0.5cm}
  \caption{The performance comparison under different probabilities of LOS blockage for user-BS links with $P = 30$ dBm and $M_{\mathsf{B}} = 12$: (left) position estimation, (right) signal detection.}\label{Fig:RMSE_LOS}
  \vspace{-0.5cm}
\end{figure}

The impact of the LOS blockage for the user-BS links on the performance of the proposed algorithm is shown in Fig. \ref{Fig:RMSE_LOS}. When the LOS blockage becomes more severe with a larger probability of LOS blockage, the multi-user tracking and signal detection performance of the proposed algorithm and Benchmark 2 is significantly degraded due to a smaller average number of position-related measurements in each tracking time slot. However, Benchmark 1 has slightly degraded performance when the probability of LOS blockage increases, since we observe that position-related measurements of user-BS links obtained by MUSIC are still unreliable and provide only mild refinement of the localization performance in the case with inadequate snapshots. Therefore,  the proposed algorithm can also make better use of the multi-path information over existing schemes enhance the tracking performance with even millimeter-level location precision.

\begin{figure}[t]
  \centering
  \includegraphics[width=.48\textwidth]{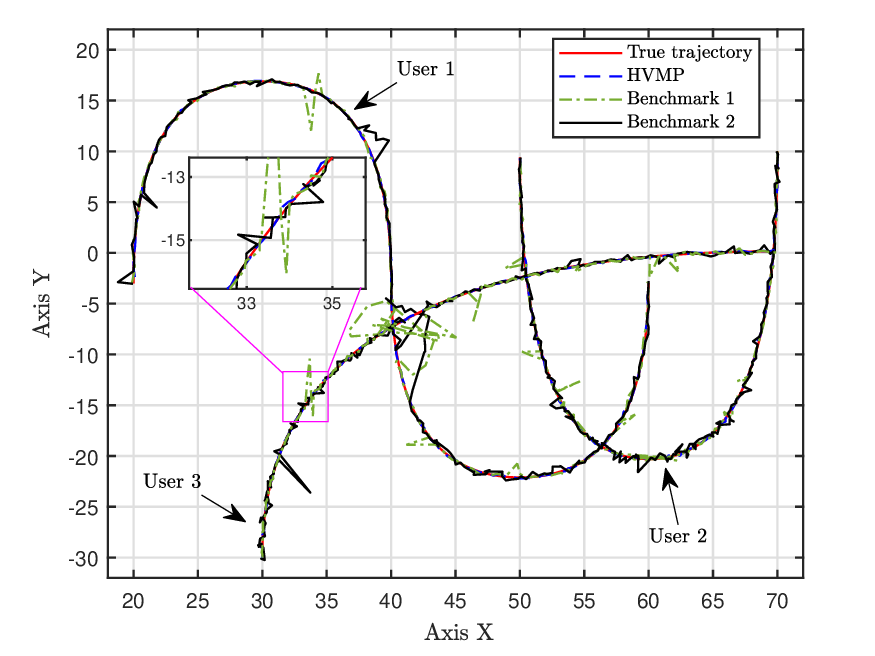}
  \vspace{-0.5cm}
  \caption{The tracked trajectory under the transmit power $ P = 10$ dBm.}\label{Fig:traj}
  \vspace{-0.3cm}
\end{figure}

\begin{figure}[t]
  \centering
  \includegraphics[width=.48\textwidth]{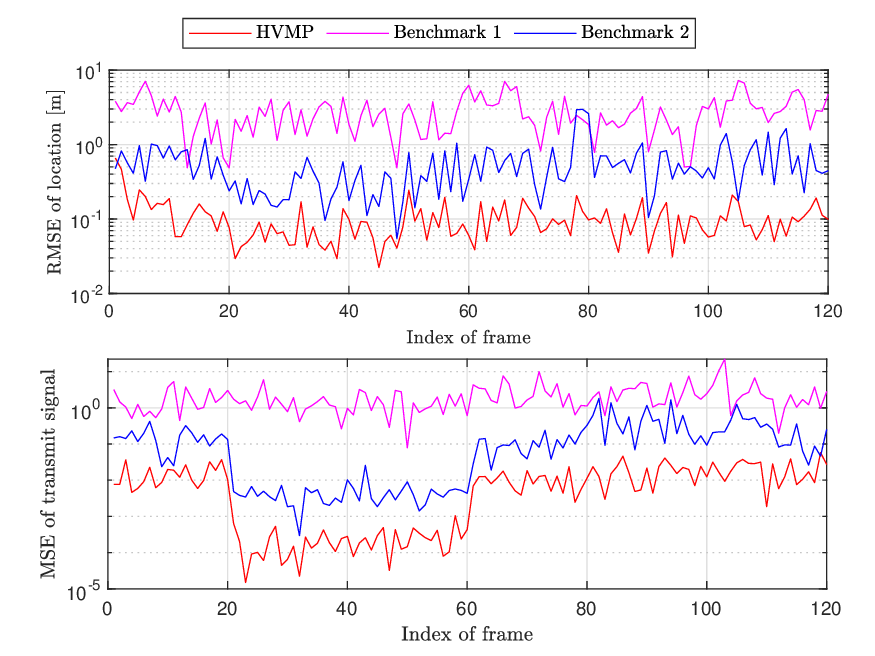}
  \vspace{-0.4cm}
  \caption{The tracking RMSE in each time slot under the transmit power $ P = 10$ dBm.}\label{Fig:RMSE_frame}
  \vspace{-0.6cm}
\end{figure}

Finally, we give an example of the tracking trajectory derived by the proposed algorithm and benchmarks in Fig. \ref{Fig:traj}. It is observed that tracked trajectories of the proposed algorithm keep approaching the true trajectories, while the two benchmarks usually fluctuate during the tracking procedure. Moreover, tracked trajectories of benchmarks have much larger deviations at the intersection region of the trajectories of different users, due to their severe mutual interferences. The tracking and signal detection performance across different time slots during the tracking procedure is also evaluated in Fig. \ref{Fig:RMSE_frame}. Here, we consider the setup where LOS links are available in the tracking time slot $t \in [21, 60]$. It is observed that the tracking performance is slightly improved with extra LOS measurements, while the signal detection performance is dramatically enhanced by exploiting the multi-path diversity. The above results also validate that the proposed algorithm can effectively combine the measurements from LOS and reflecting links to boost the joint estimation performance, making it appealing to the practical system. The future work can extend this system to the coded system by exploiting channel coding to boost the performance.


%

\section{Conclusions}

This work studied the joint multi-user tracking and signal detection problem in the RIS-assisted ISAC systems. We first established a comprehensive probabilistic model for the received signals during the tracking process and then proposed a novel HVMP algorithm with computational efficiency to simultaneously track user states (position and velocity) and detect the transmit signals in an online manner. We further analyzed the tracking performance via the performance limit of BCRB and proposed to optimize the RIS phase profile by minimizing the weighted BCRB. Simulation results were provided to validate that the proposed algorithm can realize high-accuracy tracking and signal detection performance for users with diverse and high mobility.

\begin{appendices}

\section{Derivation of the Message \eqref{equ:m_df_to_dp}}\label{sec:message_dp}

The message in \eqref{equ:m_df_to_dp} can be detailed expressed by \eqref{equ:m_df_to_dp_detail},
\begin{figure*}
\begin{small}
\begin{align}\label{equ:m_df_to_dp_detail}
  &\varepsilon_{\dot{\mathbf{p}}_{k,t} \leftarrow \dot{f}_{k,t}}(j) \varpropto  \int \prod_{g=1}^{G} \delta\left(\nu^{\mathsf{UB}}_{k,g,t} - \lambda^{-1} \zeta_{\mathsf{T}} (\dot{\mathbf{p}}^{\mathsf{U}}_{k,t})^\mathsf{T} \mathbf{e}^{\mathsf{UB}}_{k,g,t} \right)  \mathcal{VM}(-\zeta_{\mathsf{T}} \nu^{\mathsf{UB}}_{k,t}; \overleftarrow{\mu}^{\mathsf{UB}}_{k,g,t}(\nu, j), \overleftarrow{\kappa}^{\mathsf{UB}}_{k,g,t}(\nu, j))  \notag \\ 
  & \times \prod_{r=1}^{R} \delta\left(\nu^{\mathsf{UI}}_{k,r,t} - \lambda^{-1} \zeta_{\mathsf{T}} (\dot{\mathbf{p}}^{\mathsf{U}}_{k,t})^\mathsf{T} \mathbf{e}^{\mathsf{UI}}_{k,r,t} \right)  \mathcal{VM}(-\zeta_{\mathsf{T}} \nu^{\mathsf{UI}}_{k,r,t}; \overleftarrow{\mu}^{\mathsf{UI}}_{k,r,t}(\nu, j), \overleftarrow{\kappa}^{\mathsf{UI}}_{k,r,t}(\nu, j)), \notag \\
  & \varpropto \prod_{g=1}^{G} \mathcal{VM}\left(-\lambda^{-1} \zeta_{\mathsf{T}} (\dot{\mathbf{p}}^{\mathsf{U}}_{k,t})^\mathsf{T} \mathbf{e}^{\mathsf{UB}}_{k,g,t}; \overleftarrow{\mu}^{\mathsf{UB}}_{k,g,t}(\nu, j), \overleftarrow{\kappa}^{\mathsf{UB}}_{k,g,t}(\nu, j)\right)  \prod_{r=1}^{R} \mathcal{VM}\left(-\lambda^{-1} \zeta_{\mathsf{T}} (\dot{\mathbf{p}}^{\mathsf{U}}_{k,t})^\mathsf{T} \mathbf{e}^{\mathsf{UI}}_{k,r,t}; \overleftarrow{\mu}^{\mathsf{UI}}_{k,r,t}(\nu, j), \overleftarrow{\kappa}^{\mathsf{UI}}_{k,t,t}(\nu, j)\right),
\end{align}
\end{small}
\hrule
\end{figure*}
whose explicit expression is hard to obtain. By further utilizing the Gaussian approximation, the message can be obtained as \eqref{equ:m_df_to_dp}, where the mean $\dot{\overleftarrow{\mathbf{p}}}_{k,t}(j)$ and covariance $\dot{\overleftarrow{\mathbf{Q}}}_{k,t}(j)$ are obtained by the Gauss-Newton method. By denoting the exponential term in \eqref{equ:m_df_to_dp_detail} as $\mathcal{F}_{1}(\dot{\mathbf{p}}_{k,t})$, the gradient and the Hessian matrix calculated in each iteration as \eqref{equ:grad_v} and \eqref{equ:Hess_v},
\begin{figure*}
\begin{small}
\begin{align}
  \pmb{\triangledown} \mathcal{F}_1(\dot{\mathbf{p}}_{k,t})  =& - \sum_{g=1}^{G}  \frac{\breve{\zeta}_{{\mathsf{T}}} \overleftarrow{\kappa}^{\mathsf{UB}}_{k,g,t}(\nu, j)}{\lambda}  \overleftarrow{\mathbf{e}}^{\mathsf{UB}}_{k,g,t}(j) \sin\left( \frac{\breve{\zeta}_{{\mathsf{T}}}}{\lambda}  (\dot{\mathbf{p}}^{\mathsf{U}}_{k,t})^\mathsf{T} \overleftarrow{\mathbf{e}}^{\mathsf{UB}}_{k,g,t}(j) + \overleftarrow{\mu}^{\mathsf{UB}}_{k,g,t}(\nu, j)\right) \notag \\
  & - \sum_{r=1}^{R} \frac{\breve{\zeta}_{{\mathsf{T}}} \overleftarrow{\kappa}^{\mathsf{UI}}_{k,r,t}(\nu, j)}{\lambda}   \overleftarrow{\mathbf{e}}^{\mathsf{UI}}_{k,r,t}(j)  \sin\left( \frac{\breve{\zeta}_{{\mathsf{T}}}}{\lambda}  (\dot{\mathbf{p}}^{\mathsf{U}}_{k,t})^\mathsf{T} \overleftarrow{\mathbf{e}}^{\mathsf{UI}}_{k,r,t}(j) + \overleftarrow{\mu}^{\mathsf{UI}}_{k,t}(\nu, j)\right), \label{equ:grad_v} \\
  \pmb{\triangledown}^2 \mathcal{F}_1(\dot{\mathbf{p}}_{k,t}) =&  - \sum_{g=1}^{G} \frac{\breve{\zeta}_{{\mathsf{T}}}^2 \overleftarrow{\kappa}^{\mathsf{UB}}_{k,g,t}(\nu, j) }{ \lambda^{2}}  \overleftarrow{\mathbf{e}}^{\mathsf{UB}}_{k,g,t}(j) (\overleftarrow{\mathbf{e}}^{\mathsf{UB}}_{k,g,t}(j))^\mathsf{T} \cos\left( \frac{\breve{\zeta}_{{\mathsf{T}}}}{\lambda}  (\dot{\mathbf{p}}^{\mathsf{U}}_{k,t})^\mathsf{T} \overleftarrow{\mathbf{e}}^{\mathsf{UB}}_{k,g,t}(j) + \overleftarrow{\mu}^{\mathsf{UB}}_{k,g,t}(\nu, j)\right) \notag \\
  & - \sum_{g=1}^{G} \frac{\breve{\zeta}_{{\mathsf{T}}}^2 \overleftarrow{\kappa}^{\mathsf{UI}}_{k,r,t}(\nu, j)}{\lambda^{2}}   \overleftarrow{\mathbf{e}}^{\mathsf{UI}}_{k,r,t}(j) (\overleftarrow{\mathbf{e}}^{\mathsf{UI}}_{k,r,t}(j))^\mathsf{T}  \cos \left( \frac{\breve{\zeta}_{{\mathsf{T}}}}{\lambda} (\dot{\mathbf{p}}^{\mathsf{U}}_{k,t})^\mathsf{T} \overleftarrow{\mathbf{e}}^{\mathsf{UI}}_{k,r,t}(j) + \overleftarrow{\mu}^{\mathsf{UI}}_{k,r,t}(\nu, j) \right), \label{equ:Hess_v}
\end{align}
\end{small}
\hrule
\end{figure*}
where $\overleftarrow{\mathbf{e}}^{\mathsf{UB}}_{k,g,t}(j) = (\overleftarrow{d}^{\mathsf{UB}}_{k,g,t}(j))^{-1} (\overleftarrow{\mathbf{p}}^{\mathsf{UB}}_{k,g,t}(j) - \mathbf{p}^{\mathsf{B}}_{g})$ is the updated directional vector for the user $k$-BS $g$ link in the $j$th iteration of HVMP and $\overleftarrow{d}^{\mathsf{UB}}_{k,g,t}(j)$ is the updated distance for the user $k$-BS $g$ link with $\overleftarrow{\mathbf{p}}^{\mathsf{UB}}_{k,g,t}(j)$ updated in \eqref{equ:m_pUB_to_f}.
Then, the updated value of $\dot{\overleftarrow{\mathbf{p}}}_{k,t}(j,\iota_{\rm vel})$ and $\dot{\overleftarrow{\mathbf{Q}}}_{k,t}(j,\iota_{\rm vel})$ in each iteration can be given by
\begin{align}
  &\dot{\overleftarrow{\mathbf{p}}}_{k,t}(j,\mathsf{new})  = \dot{\overleftarrow{\mathbf{p}}}_{k,t}(j,\mathsf{old}) \notag \\
  &- \left(\pmb{\triangledown}^2 \mathcal{F}_{1}(\dot{\overleftarrow{\mathbf{p}}}_{k,t}(j,\mathsf{old}))\right)^{-1} \pmb{\triangledown} \mathcal{F}_{1}(\dot{\overleftarrow{\mathbf{p}}}_{k,t}(j,\mathsf{old})), \\
  &\dot{\overleftarrow{\mathbf{Q}}}_{k,t}(j,\mathsf{new}) = -\left(\pmb{\triangledown}^2 \mathcal{F}_{1}(\dot{\overleftarrow{\mathbf{p}}}_{k,t}(j,\mathsf{new})\right)^{-1}.
\end{align}
After the Gauss-Newton method converges, we can finally obtain the mean and covariance in the Gaussian message \eqref{equ:m_df_to_dp}.

\end{appendices}

\bibliographystyle{IEEEtran}
\bibliography{ref}

\end{document}